\documentclass[usenatbib]{mn2e}

\usepackage[dvips]{graphicx,color}
\usepackage{aas_macros}

\usepackage{amsmath}
\usepackage{pifont}
\usepackage{lscape}

\newcommand\hii{UC H\,{\sc ii}}
\newcommand\mdot{M$_{\odot}$}

\begin{document}

\title[Near- and Mid-Infrared Spectroscopy of Young Massive Stars in W31]
{Mid-Infrared diagnostics of metal-rich H\,{\sc ii} regions from VLT and 
{\it Spitzer}
Spectroscopy of Young Massive Stars in W31\thanks{Based on observations
made with ESO telescopes at the Paranal Observatory under programme
ID 077.C-0550(A) and the Spitzer Space Telescope which is operated
by the Jet Propulsion Laboratory, California Institute of Technology
under NASA contract 1407}}
\author[J. P. Furness et al.]{J. P. Furness$^{1}$, P. A. 
Crowther$^{1}$\thanks{Paul.Crowther@shef.ac.uk},  P. W. Morris$^{2}$, C. 
L. Barbosa$^{3}$, R. D. Blum$^4$, \newauthor P. S. Conti$^{5}$, S. D.
van Dyk$^6$ \vspace{3mm} \\ 
$^{1}$Department of Physics and Astronomy, University of Sheffield, 
Sheffield S3 7RH, UK\\
$^{2}${NASA Herschel Science Center}/CalTech, 220-6, Pasadena, CA 91125, USA\\
$^{3}$ IP\&D, Universidade do vale do Para\'{\i}ba, Av. Shishima Hifumi, 
2911,S\~ao Jos\'e dos Campos 12244-000, SP, Brazil\\
$^{4}$NOAO, 950 North Cherry Avenue, Tucson, AZ 85719, USA \\
$^{5}$JILA, University of Colorado, Boulder, CO 80309-0440, USA\\
$^{6}$Spitzer Science Center/CalTech, 220-6, Pasadena, CA 91125, USA}
\date{\today}

\pagerange{\pageref{firstpage}--\pageref{lastpage}} \pubyear{2008}

\maketitle

\label{firstpage}

\begin{abstract}
We present near-IR VLT/ISAAC and mid-IR {\it Spitzer}/IRS spectroscopy
of the young massive cluster in the W31 star-forming
region. $H$--band spectroscopy provides refined classifications
for four cluster members O stars with respect to Blum et al.
\nocite{Blum01} In addition, photospheric features
are detected in the massive Young Stellar Object (mYSO) \#26. 
Spectroscopy permits estimates of stellar temperatures and masses, from 
which a cluster age of $\sim$0.6 Myr and distance of 3.3\,kpc are obtained, 
in excellent agreement with Blum et al\nocite{Blum01}. IRS spectroscopy 
reveals mid-infrared fine structure line fluxes of [Ne\,{\sc ii-iii}] and
[S\,{\sc iii-iv}] for four O stars and five mYSOs. In common with previous studies,  
stellar temperatures of individual stars are severely underestimated from
the observed ratios of fine-structure lines, despite the use of contemporary
stellar atmosphere and photoionization models.
We construct empirical temperature calibrations based upon the W31 cluster stars
of known spectral type, supplemented by
two inner Milky Way ultracompact (UC) H\,{\sc ii} regions whose 
ionizing star properties are
established. Calibrations involving [Ne\,{\sc iii}] 15.5$\mu$m/[Ne\,{\sc ii}]
12.8$\mu$m, [S\,{\sc iv}] 10.5$\mu$m/[Ne\,{\sc ii}] 12.8$\mu$m or 
[Ar\,{\sc iii}] 9.0$\mu$m/[Ne\,{\sc ii}] 12.8$\mu$m have 
application in deducing the spectral types of early- to mid- O stars for other 
inner Milky Way compact and \hii\ regions.
Finally, evolutionary phases and timescales for the massive stellar 
content in W31 are discussed, due to the presence of numerous young 
massive stars at different formation phases in a `coeval' cluster. 
\end{abstract} 
%
%
% A similar approach
% for compact H\,{\sc ii} regions beyond the Solar circle is also carried 
% out. 
% 

%\vspace{0.5cm}


\begin{figure*}
\includegraphics[width=\columnwidth,clip]{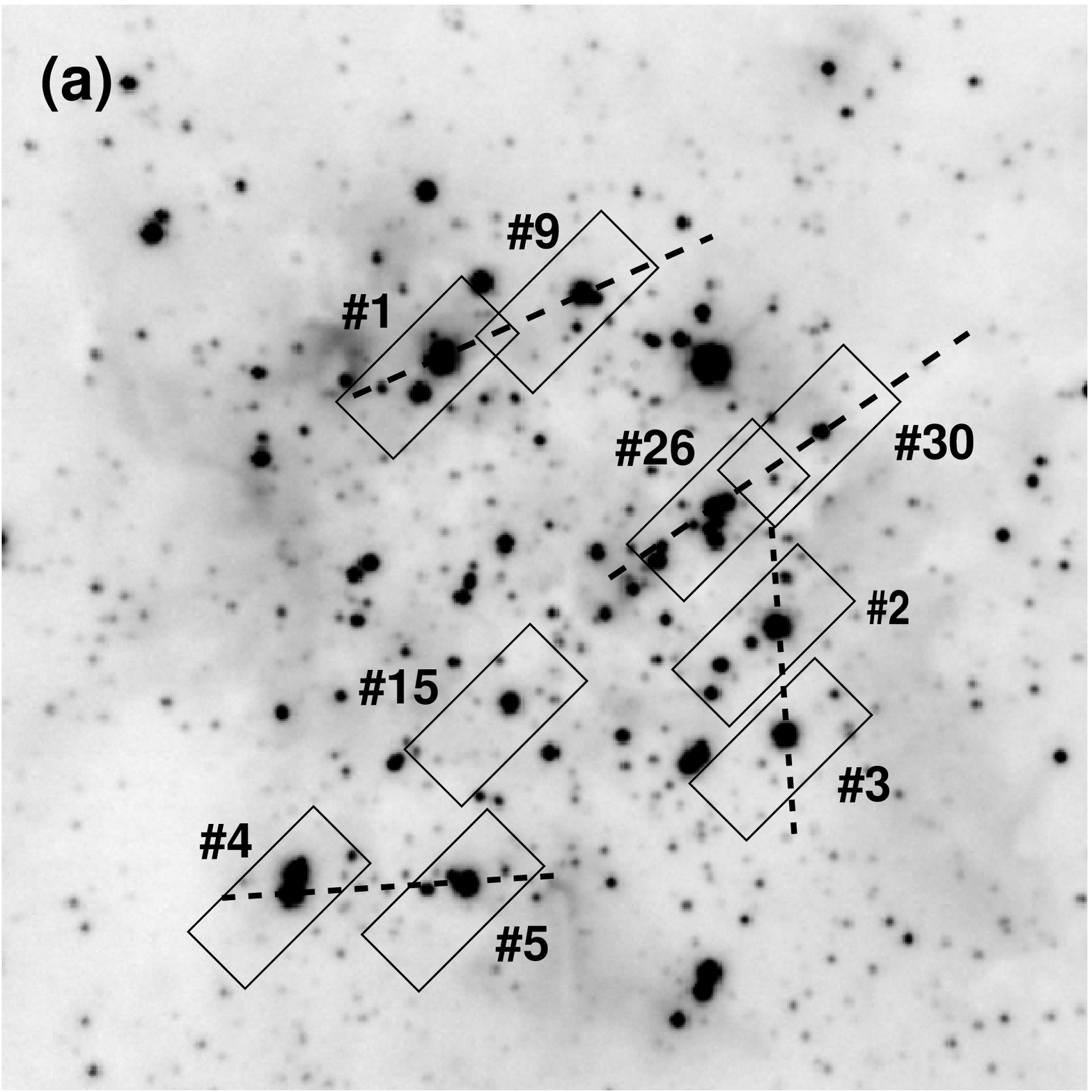}
\includegraphics[width=\columnwidth,clip]{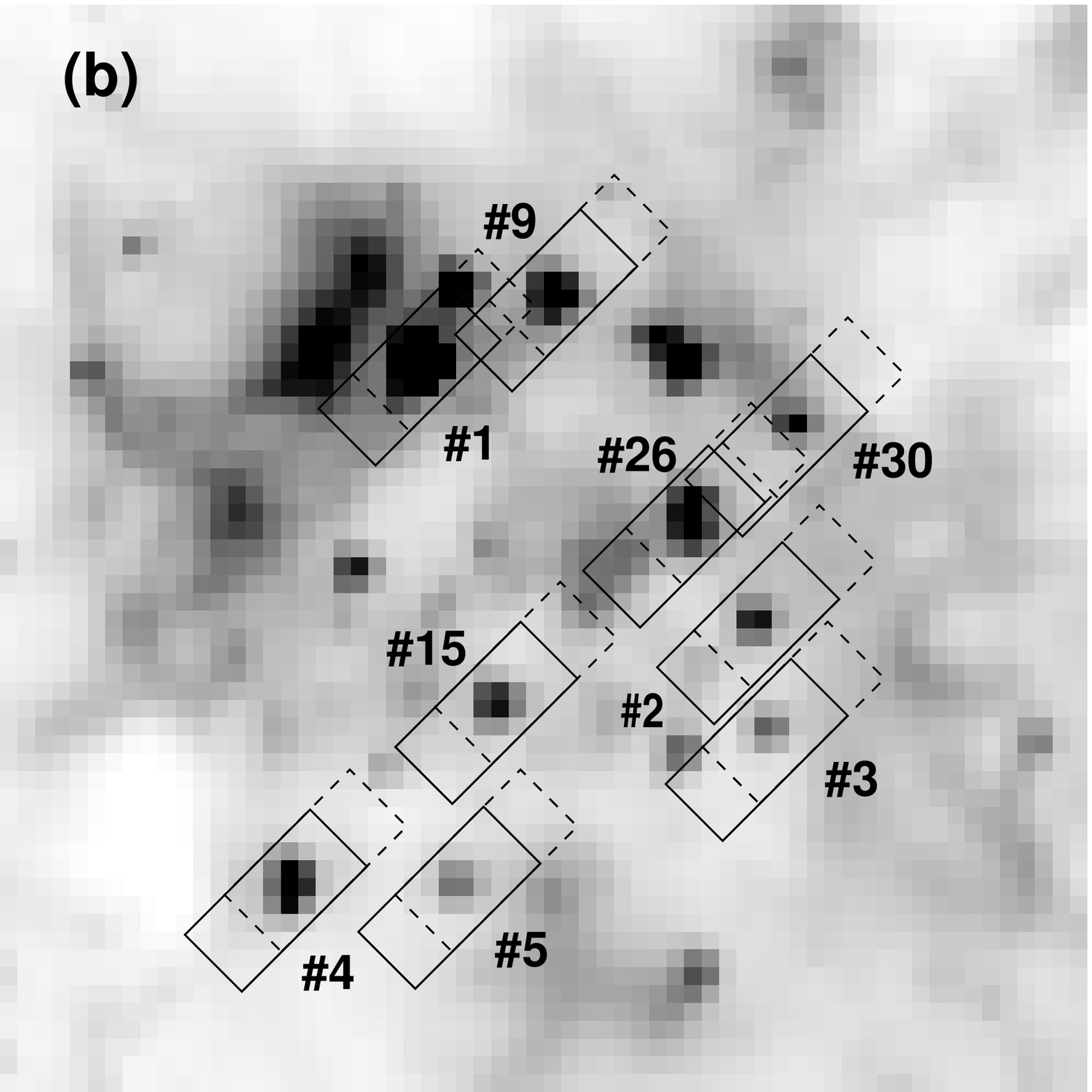}
\caption{(a) $K$-band image of W31 from \citet{Blum01}, showing 
the 4 slit positions for the near-IR ISAAC spectroscopy (dashed lines) and
the 9 IRS apertures for the first nod position of the
mid-IR spectroscopy. North is up, East to
the left and the image is 72$''$ $\times$ 72$''$ or 1.2 $\times$ 1.2 pc at 
our adopted distance of 3.3 kpc;  
(b) 3.6$\mu$m IRAC image of W31 showing both IRS
nod positions (first solid; second dotted) at the same 
scale and orientation as for (a).}
\label{slits}
\end{figure*}

\section{Introduction}

The formation of high mass stars remains an unsolved astrophysical puzzle
\citep{Zinnecker07, Clarke08}. Unlike the situation for low mass 
stars, for which
multiwavelength observations are plentiful, 
very massive stars ($M_{\ast} \geq$25\mdot) 
are born within compact, deeply embedded star-forming regions, severely
restricting observations to either 
the far-infrared where heated dust dominates their appearance, or
radio wavelengths where gas ionized by the central star(s) can be 
characterised. Either way, high mass stars themselves cannot be seen directly
within these ultracompact (UC) H{\sc ii} regions \citep{Church02} 
until the column density of dust along our line of sight falls below 2--4 
magnitudes in the $K$-band. 
To date, only a few such cases have
been identified -- G23.96+0.15 \citep*{Hanson02, Crowther08},
G29.96--0.02 \citep{Watson, Hanson227}, 
G45.45+0.06 \citep{Blum08}, W51d \citep{Barbosa08} -- with
the stellar content of other \hii\ regions reliant upon indirect far-IR
or radio continuum techniques \citep{WC89, Kurtz94}.

Fortunately, the advent of efficient mid-infrared imaging and spectroscopy
from space with {\it Infrared Space Observatory} 
\citep[{\it ISO},][]{Kessler96} 
and
{\it Spitzer} \citep{Werner04}, plus 
ground-based 8--10m telescopes has opened up a further window to 
study such embedded  regions. Specifically, a number of fine-structure
lines from ionized regions are seen in the mid-infrared, 
notably [Ne\,{\sc ii-iii}], [S\,{\sc iii-iv}], which albeit 
indirect, provide information upon the hardness of the extreme 
ultraviolet (EUV) radiation from their constituent O stars. Such diagnostics,
analogous to the optical forbidden lines of [O\,{\sc ii-iii}] 
and [S\,{\sc ii-iii}], may allow the `inverse problem' of establishing 
properties of the ionizing stars of ultra-compact and compact 
H\,{\sc ii} regions \citep[see][]{Okamoto03}. Alternatively,
indirect approaches based upon near-IR hydrogen and helium nebular
lines are also employed \citep[e.g.][]{Lumsden, Blum09}.

To date, tests of photoionization and stellar atmosphere models using
these mid-IR line diagnostics have been rather unsatisfactory. Firstly,
this is because H\,{\sc ii} regions are usually ionized by multiple
early-type stars in compact clusters, and secondly the strength of 
mid-IR fine structure lines is affected both by the energetic
photons from OB stars and nebular properties. The most comprehensive study 
of a \hii\ region attempted to date has been by \citet{Morisset02} for 
G29.96--0.02 whose results suggested a temperature of $T_{\rm 
eff} \sim35\pm3$kK for
the ionizing star, in contrast to $T_{\rm eff} \sim41\pm2$kK from a 
non-LTE
analysis of near-IR spectroscopy for the star 
\citep*{Hanson227}.

Photoionization models are widely used to infer the stellar properties of 
both embedded H\,{\sc ii} regions \citep{Sellmaier96, Giveon02, 
SimonDiaz08, PerezMontero09} and entire galaxies from mid-IR fine 
structure lines 
\citep{Lutz96, Rigby04}. If  the highly discrepant effective temperatures 
obtained for G29.96--0.02 using indirect techniques were 
repeated for other, single, embedded O stars, then
previously published results from mid-IR
diagnostics may be called into question, especially
those at high metallicity \citep[e.g.][]{Thornley00}.

The focus of the present study is the Galactic Giant H{\small II} region 
(GHR) W31, specifically the young star cluster (10.2--0.3, hereafter W31) 
discussed by \citet*{Blum01}. This cluster, for which \cite{Blum01} 
established a distance of 3.4 kpc and extinction of $A_{\rm K}$ = 1.7 mag, 
hosts a minimum of four `naked' O-type stars, as deduced from near-IR 
spectroscopy, plus a number of massive stars which are still embedded at 
near-IR wavelengths (hereafter massive YSOs) plus numerous \hii\ regions 
\citep{Ghosh}. As such, this cluster provides an excellent opportunity to 
study the different early evolutionary phases of massive star formation. 
In addition, it possesses an unusual morphology, in that the highest mass 
stars are located at the periphery of the cluster, in contrast to the 
mass segregated morphology of most other young clusters 
\citep{degrijs02,
allison09}.
We shall exploit 
this unusual geometry through spectroscopy 
of individual early-type stars in W31 with the Infrared 
Spectrograph\citep[IRS,][]{SpitzerIRS} aboard {\it Spitzer}, 
supplemented with $H$ and $K$ band VLT observations with
the Infrared  Spectrograph and Array Camera\citep[ISAAC,][]{ISAAC}.

\begin{figure*}
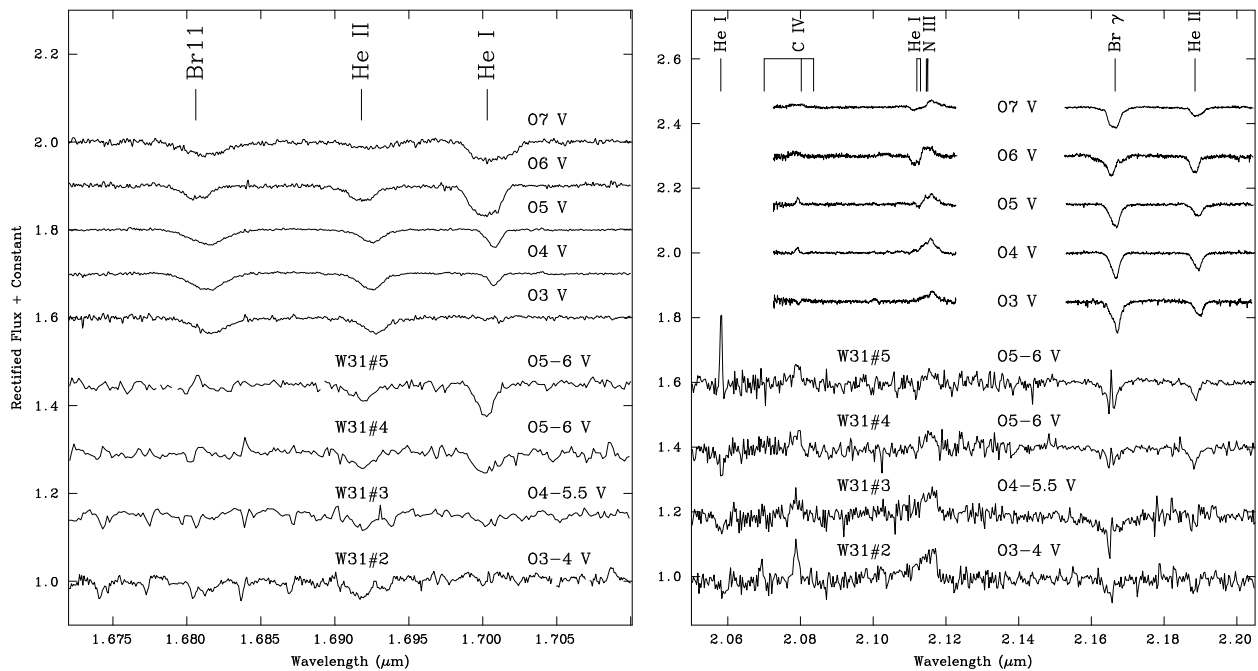

\includegraphics[width=0.5\textwidth,angle=270,clip]{ostars_h.eps}\includegraphics[width=0.5\textwidth,angle=270,clip]{ostars_k.eps}
\caption{$H$ band (left panel) and $K$ band (right panel) VLT/ISAAC
spectra of
the O stars in W31, plus template stars from \citet{Hanson05}.  
All stars show stellar absorption of He\,{\sc ii} and He\,{\sc 
i} in the $H$ band, the ratio of which we use to spectral type the
stars. The stars also show C\,{\sc iv} and N\,{\sc iii} emission in the $K$-band indicating early--mid
type O stars. The later of the O stars also show Br$\gamma$ emission,
indicating a younger evolutionary phase where more nebula material is
present around the star.}
\label{ostars}
\end{figure*}

Our aim is to compare photoionization model predictions of individual 
H\,{\sc ii} regions with stellar temperatures obtained directly from 
near-IR spectroscopic analysis. We are thus able to significantly increase 
the statistics of H\,{\sc ii} regions for which mid-IR diagnostics are 
available, whose ionizing stars are well constrained. Our results 
are of 
potential application in deducing ionizing stars of \hii\ regions within 
the inner Milky Way based upon fine structure diagnostics available from 
ground-based telescopes, namely [S\,{\sc iv}] 10.5$\mu$m and [Ne\,{\sc 
ii}] 12.8$\mu$m \citep{Zhu08}. The diffraction limit of 8--10m telescopes 
in the N-band is $\sim$0.4 arcsec, equivalent to a physical scale of 0.01 
pc at a distance of 5 kpc, well below the 0.1 pc size of typical \hii\ 
regions \citep{Okamoto03}.

The present paper is structured as follows.
\S~\ref{obs} detail the VLT and {\it Spitzer} observations, while \S 
\ref{3p1} 
presents refined
spectral classification of naked O stars  in W31. 
\S~\ref{finestruc} presents the mid-IR fine structure lines from 
W31 sources, including predictions from photoionization models and 
empirical calibrations. A discussion of mid-IR fine structure lines is 
presented \S~\ref{discussion} together with the massive star content of 
W31. Brief conclusions are drawn in \S~\ref{summary}.

\section{Observations and Data Reduction} \label{obs}

Two main observational datasets are used in the present study, obtained
with the VLT ISAAC near-IR spectrograph and {\it Spitzer} IRS mid-IR 
spectrograph.

\subsection{VLT ISAAC spectroscopy}

Long-slit $H$- and $K$-band near-infrared spectroscopy of sources in W31 were obtained
with the ISAAC instrument mounted at the Very Large Telescope between 
4th April--17th June 2006 (Programme 077.C-0550(A), P.I. Crowther).  
Four slit positions were used,  as illustrated in Fig.~\ref{slits}(a). 
Each included two or more targets of interest, 
namely the naked O stars and massive YSO's from \cite{Blum01}.

The detector was the 1024$\times$1024 Hawaii Rockwell array, while three 
medium resolution grating settings (0.775\AA/pixel) were obtained, centred 
at 1.71, 2.09 and 2.20\,$\mu$m. These observations were obtained at 
low airmass during variable seeing conditions using a 0.6 arcsec wide 
slit, and reduced  using standard \textsc{iraf} packages. We observed 
using ABBA nod-cycles,  a standard infrared A number of AB pairs were 
obtained for each grating 
setting with wavelength solutions achieved from comparison XeAr arc 
images. From these, the observations covered 1.671--1.751\,$\mu$m, 
2.029--2.155\,$\mu$m and 2.140--2.265\,$\mu$m at spectral resolutions of 
3.8\AA, 6.0\AA\ and 6.0\AA\ respectively, as measured from arc lines.

\begin{table*}
\begin{center}
\caption{Classification of early-type stars in W31 from VLT/ISAAC 
spectroscopy based upon the near-IR calibration of \citet{Crowther08}.}
\begin{tabular}{lccrlll}
\hline 
               &  \multicolumn{2}{c}{EW (\AA)}   &      &   \multicolumn{2}{c}{Spectral Type}  &                                \\
               &  He\,{\sc ii} (1.692$\mu$m)  & He\,{\sc i}(1.700$\mu$m)  
& log$\frac{\rm He\,II}{\rm He\,I}$  &  
This Work & Blum et al.  &  Further Comments   \\ \hline
%Hanson Atlas   &                     &                   &    &        &         &               \\
%HD64568   &  0.718 $\pm$ 0.016  & 0.041 $\pm$ 0.006 &  17.5   &  O3V   &         &                \\ 
%HD46223   &  0.686 $\pm$ 0.008  & 0.231 $\pm$ 0.005 &  2.97   &  O4V   &         &                \\
%HD46150   &  0.549 $\pm$ 0.007  & 0.441 $\pm$ 0.005 &  1.25   &  O5V   &         &                \\ 
%HD5689    &  0.795 $\pm$ 0.018  & 1.657 $\pm$ 0.016 &  0.48   &  O6V   &         &                \\ \vspace{4mm}
%HD217086  &  0.476 $\pm$ 0.024  & 1.365 $\pm$ 0.022 &  0.35   &  O7V   &         &                \\ 
%Observed Sources &              &                   &         &        &         &                \\
 \#2   &  0.59 $\pm$ 0.08  & $<$ 0.2         & $>$0.47$\pm$0.06   &  O3--4\,V   &  O5.5\,V  &  N\,{\sc iii}, C\,{\sc iv} in emission \\  
 \#3   &  0.40 $\pm$ 0.08  & 0.27 $\pm$ 0.07 & 0.17$\pm$0.20   &  O5$^{+0.5}_{-1}$\,V   &  O5.5\,V  &  N\,{\sc iii}, C\,{\sc iv} in emission \\
 \#4   &  0.58 $\pm$ 0.06  & 0.87 $\pm$ 0.06 & --0.17$\pm$0.07  &  O5.5$\pm$0.5\,V &  
O5.5V  &  N\,{\sc iii}, C\,{\sc iv} in emission \\  
 \#5   &  0.68 $\pm$ 0.05  & 0.96 $\pm$ 0.04 & --0.15$\pm$0.05  &  O5.5$\pm$0.5\,V &  
O5.5V  &  N\,{\sc iii}, C\,{\sc iv} in emission, nebular \\
          &                     &                   &         &  &   & emission from Br$\gamma$, He\,{\sc i} 2.058$\mu$m\\
% W31 \#26  &  0.444 $\pm$ 0.071  & 0.806 $\pm$ 0.061 & -0.260  &  O6V   &  MYSO   &  NIII in emission\\
\hline
\end{tabular} 
\label{sptypes}
\end{center}
\end{table*}
%Flux ratio $\frac{{\rm He II} (\lambda 16918 \AA)}{{\rm He I} (\lambda 17003 \AA)}$ & Spectral Type

Telluric correction was achieved by spectroscopy of early-G dwarfs 
observed at similar airmass to W31, corrected for their spectral features 
using high resolution observations of the Sun, adjusted for the radial 
velocity and spectral resolution of the template stars. An 
extensive discussion of telluric correction for medium 
resolution near-IR spectroscopy of early-type stars 
is provided by \citet{Hanson05}. The 2.09$\mu$m 
setup suffered from low-level variable structure which was accentuated 
upon flat-fielding, arising from a 50 Hz pickup inherent to the 
instrument. Only the two other settings were flat-fielded. Consequently, 
the continuum S/N achieved was lower for this grating position ($\sim$100) 
than the other settings ($\sim$200).

\subsection{{\it Spitzer} IRS spectroscopy}\label{sect2.2}

W31 was observed with the mid-IR spectrograph IRS in GO \#3337 (W31CLUST, 
P.I. Crowther) between 14-16 September 2005 using all four modules, 
sampling the short wavelength region at low and high resolution (SL and 
SH), plus the long wavelength region at low and high resolution (LL and 
LH). In addition, W31 was imaged with the IRAC instrument 
\citep{Fazio04}
at 3.6, 4.8, 5.8 
and 8.0$\mu$m, using HDR mode, with 12 s exposures in a 12 position, 
Reuleaux dither pattern. Unfortunately, the IRAC 5.8 and 8.0$\mu$m
imaging and long 
wavelength IRS observations were heavily saturated, so our analysis largely
focuses upon the SH, staring mode (30 sec, 6 cycles) 
observation with IRS, obtained 
at two nod positions. The spectral 
range covered was 9.9--19.6{\rm $\mu$}m, at a 
resolution of R$\sim$600.

The individual 4.7$\times$11.3 arcsec$^{2}$ (2$\times$5 pix) apertures from
the first nod position are superimposed upon a $K-$band image of W31 
in the Fig~\ref{slits}(a). Fig~\ref{slits}(b)
shows both IRS nod positions together with the IRAC 3.6$\mu$m image of W31.
A background subtraction was applied to these two dimensional IRS datasets -- prior
to extraction -- using a dedicated OFF position at 18:09:30.4 --20:20:00 (J2000)
from a second set of cluster positions which are not presented in this
paper. This pointing gives a correction for nebular emission in
the extended H\,{\sc ii} region, including the PAH's. However, it does
not correct for emission from the dense material close to each of the
targeted stars.

Observations have been processed and calibrated in the standard IRS
pipeline to basic calibrated data products (BCDs), using a version which
is compatible with S16.2. Multiple exposures at the same position have been
co-added at the BCD (2 dimensional) level, and the OFF position has been used to correct 
rogue pixels and subtract the extended background. Then the spectra were 
extracted with the offline S16.2 post-BCD pipeline. Spectra from the two 
staring-mode nods were averaged, except for cases for which a single nod was used
to minimise contamination. Specifically, only nod 1 was used for \#1 and \#26
and only nod 2 was used for \#2 and \#30.
This is to avoid contamination from potentially interloping sources during
extraction of the spectra (recall Fig.~\ref{slits}(b)). 
These sources are identified within the nods from 
any of the available imaging observations, including unsaturated regions of
the IRAC datasets.

Since the width of the full width at
half maximum of the central Airy pattern of the IRS point source function
is comparable to the size of the SH slit itself, the full slit is 
extracted,  so it is not possible to extract (or even easily identify) 
separate sources or the background.

Each of the echelle spectral orders were trimmed at the red edges,
corresponding to the lower part of the array where photometric response sharply drops.
Final spectra have been created by merging the orders into single spectra,
with excellent inter-order agreement when the point-source flux calibration
is applied, with only minor offsets (well within the photometric uncertainties
of 15\%), indicating that the dominant source signal from each of the intended
targets behave more consistently as point sources than extended sources. 
Formally, a 1$\sigma$ absolute flux calibration 
of 10\% was achieved, except for the raw 18.7{\rm $\mu$}m [S\,{\sc iii}] 
line which was saturated in a few cases. These were corrected by 
linearization and extrapolation, from which an additional 5\% uncertainty 
resulted.

Finally, we should
note that these observations have been obtained when the SH array was relatively
undamaged by the cumulative effects of space weather, thus receive good
correction for rogue pixels with data at the OFF position. Furthermore, these
observations are not affected in the calibrations or data quality by more recent
versions of the pipelines (through S18.x) which mainly benefit the IRS
peak up imaging AOT and data obtained with more aged arrays.

% Both dust and gas are expected to be highly spatially variable across W31. 
% although no background correction was possible for the IRS spectroscopy.

% In addition long wavelength high resolution (LH) observations were
% obtained but due to the faint dust continuum, emission lines were
% strongly saturated, preventing measurements of the [NeIII] (36.1{\rm
% $\mu$}m) or [SIII] (33.5{\rm $\mu$}m) fine structure
% lines. Measurements of both [NeIII] lines (15.55 and 36.1\micron)
% would have allowed a determination of the nebular density of W31.

\section{Properties of early-type stars in W31 from near-IR spectroscopy}\label{3p1}

\subsection{Near-IR classification of W31 O stars}

In Figure \ref{ostars} we present our $H$- and $K$-band ISAAC spectroscopy 
of naked O stars in W31, together with spectra of template O stars from 
the high resolution atlas of \citet{Hanson05}. \cite{Crowther08} have 
recently presented a method of determining O dwarf subtypes from the ratio 
of the observed He\,{\sc ii} 1.692$\mu$m to He\,{\sc i} 1.700$\mu$m 
equivalent widths. These were measured from the emission line fitting 
(ELF) suite of routines in the \textsc{starlink} spectroscopic analysis 
programme \textsc{dipso} and are presented in Table \ref{sptypes} together 
with inferred spectral types.

Stars \#3, \#4 and \#5 all show clear He\,{\sc i} and He\,{\sc ii} 
absorption features, from which O5$^{+0.5}_{-1}$\,V, O5.5$\pm$0.5\,V and O5.5$\pm$0.5\,V 
classifications are 
obtained, according to Fig.~3 of  \citet{Crowther08}, 
in close agreement with \citet{Blum01}. $K$-band spectroscopy 
reveals C\,{\sc iv}, N\,{\sc iii} emission features, plus Br$\gamma$ and 
He\,{\sc ii} 2.189$\mu$m absorption, the core of the former filled-in by 
nebular Br$\gamma$ emission.

In the case of \#2, negligible He\,{\sc i} 1.700$\mu$m is observed, 
from 
which an O3--4\,V classification is inferred. If we had classified \#2 on the 
basis of its $K$-band spectrum, the presence of significant C\,{\sc iv} 
emission would  have suggested a spectral type of O4V or later 
\citep{Hanson05}.

\begin{table}
\caption{Interstellar extinctions, $A_{\rm K}$, and distance modulus (DM) towards
individual stars in W31 using near-IR photometry of \citet{Blum01} and the
absolute magnitude calibration of \citet{Martins06}}
\begin{tabular}{l@{\hspace{-2mm}}c
@{\hspace{2mm}}c@{\hspace{2mm}}c@{\hspace{2mm}}c@{\hspace{2mm}}c@{\hspace{2mm}}c
@{\hspace{2mm}}c}
\hline
W31       &  $m_K$  & $H - K$  & $(H - K)_{0}$ & $E_{\rm H-K}$ & $A_K$ &   
M$_K$      &  $DM$             \\ 
   \#      & mag     & mag      & mag           & mag           & mag & 
mag \\
\hline
2          & 10.02  &  1.16 & --0.10 & 1.26 & 2.29  & -4.98  & 
12.71\\ 
3          & 10.30  &  1.06 & --0.10 & 1.16 & 2.11  & -4.39  & 
12.58\\  
4          & 10.34   &  0.88  & --0.10 & 0.98 & 1.78  & -4.27  & 12.83\\
5          & 10.37  &  1.26  & --0.10 & 1.36 & 2.47  & -4.27 & 12.17\\ 
%\vspace{1mm} 
% \#26         & 11.51 $\pm$ 0.03  &  2.72 $\pm$ 0.04 & (5.04 $\pm$ 0.16) & -4.42 $\pm$ 0.5 & (10.60 $\pm$ 0.03)  \\ \hline
Average      &                   &           &       & & 2.16        &            
& 12.57 \\ 
$\sigma$ &                      &            &        & & 0.29 && 0.29 \\
\hline
\end{tabular}  
\label{photom}
%\flushleft{Parentheses indicate source with circumstellar material,
%and the distance modulus is calculated excluding the IR excess}
\end{table}

\begin{figure}
\includegraphics[angle=270, width=\columnwidth,clip]{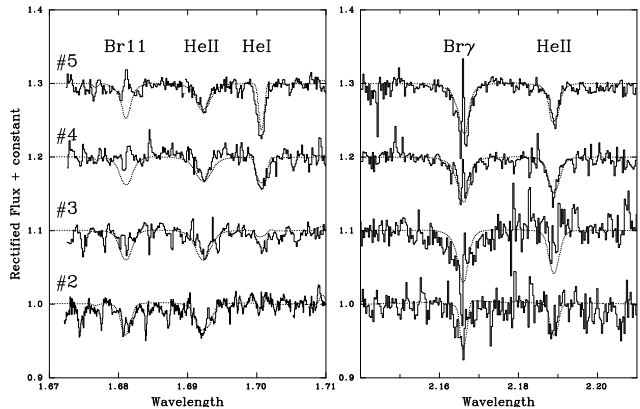}
\caption{$H$-band and $K$-band spectroscopic fits (dotted lines) to VLT/ISAAC (solid lines)
observations of hydrogen and helium lines in W31 \#2 (O3--4\,V), \#3 (O5$^{+0.5}_{-1}$\,V), \#4 (O5.5$\pm$0.5\,V) 
and \#5 (O5.5$\pm$0.5\,V).}
  \label{fits}
\end{figure}

\begin{table}
\begin{center}
\caption{Stellar properties of the W31 O stars from non-LTE CMFGEN models
using a distance of 3.3 kpc. Uncertainties in stellar temperatures
are $\pm$1.5 kK, with the exception of \#2 for which an upper limit is 48\,kK.
current stellar masses are obtained from comparison with 2$\times 
Z_{\odot}$  theoretical isochrones from \citet{Lejeune01}.}
\begin{tabular}{l
@{\hspace{2mm}}
l@{\hspace{2mm}}
l@{\hspace{2mm}}
l@{\hspace{2mm}}
l@{\hspace{2mm}}
l@{\hspace{2mm}}
l@{\hspace{0.5mm}}
l@{\hspace{2mm}}l}
\hline
W31       & Sp. & T$_{\rm eff}$ & $\log L$ & $\log \dot{M}$ & log g & 
$v \sin i$  &  M& $M_{\rm K}$ \\
\#             & Type & kK           & $L_{\odot}$ & $M_{\odot}$yr$^{-1}$ 
& cgs 
 & km\,s$^{-1}$ & $M_{\odot}$ & mag\\
\hline
2    & O3--4\,V      & 45  &  5.79    &  --5.5 & 4.00 & 100 &  61 &--4.84 \\
3    & O5$^{+0.5}_{-1}$\,V      & 43  &  5.56    &  --5.8 & 4.04 & 200 & 46  & --4.38 \\ 
4    & O5.5$\pm$0.5\,V    & 41  &  5.36    &  --6.1 & 4.04 & 200 & 36 & --4.01 \\
5    & O5.5$\pm$0.5\,V    & 41  &  5.62    &  --5.9 & 3.94 & 100 & 45 & 
--4.67 \\
%\#26         & O6V      & 36$\pm$2      & 3.92  & 5.39                       &  30$\pm$5	  \\
\hline
\end{tabular}  
\label{stellar}
\end{center}
\end{table}

\subsection{Distance and Age of W31}\label{distances}

Armed with our refined spectral types, we now use the near-IR photometry 
from \cite{Blum01} and observational absolute magnitude calibration of 
\cite{Martins06} to obtain a revised distance to W31. This is presented 
in Table~\ref{photom}, in which observed colours of 
O stars provide a direct measurement of interstellar extinction from 
$A_{K} \sim 1.82 E_{H-K}$ from \cite{Indebetouw05}. 
For \#4 and \#5 uncertainties in subtypes have little influence upon 
absolute magnitudes, while subtype uncertainties formally introduce
absolute magnitude uncertainties of up to 
0.3 mag for \#2 and \#3. These are mitigated somewhat by the 
typical spread of $\pm$0.5 mag in absolute magnitudes for individual 
spectral types.

Overall, we obtain a higher interstellar extinction of $A_{K}$ = 2.16 $\pm$0.29
mag towards W31 than \cite{Blum01} on the basis of updated intrinsic colours, 
but this is largely cancelled out by the revised absolute magnitude calibration.
As such, we obtain a very similar overall distance (3.3 $_{-0.3}^{+0.4}$ kpc)
to the 3.4 kpc distance obtained by \cite{Blum01}.

For an adopted Galactic Centre distance of 8.0 kpc \citep{Reid93} for the 
Sun, our preferred distance to  W31 suggests that it
lies $\sim$4.8 kpc from the Galactic Centre.
The Galactic oxygen metallicity gradient is $\Delta \log$ (O/H) = --0.044
$\pm$ 0.010 dex\,kpc$^{-1}$ \citep{Esteban05}, from which we anticipate
that W31 is 40\% more metal-rich than H\,{\sc ii} regions 
within the Solar circle, i.e. $\log$ O/H +12 $\sim$ 8.81, compared to
8.65 for the  Orion Nebula at a Galactocentric distance of $\sim$8.4 kpc
\citep{Esteban05, SimonDiaz06}. We are unable to derive a neon or
sulphur abundance for W31 from the IRS spectroscopy since the nebular electron
temperature is unknown.

\begin{figure}
\includegraphics[angle=0, width=\columnwidth,clip]{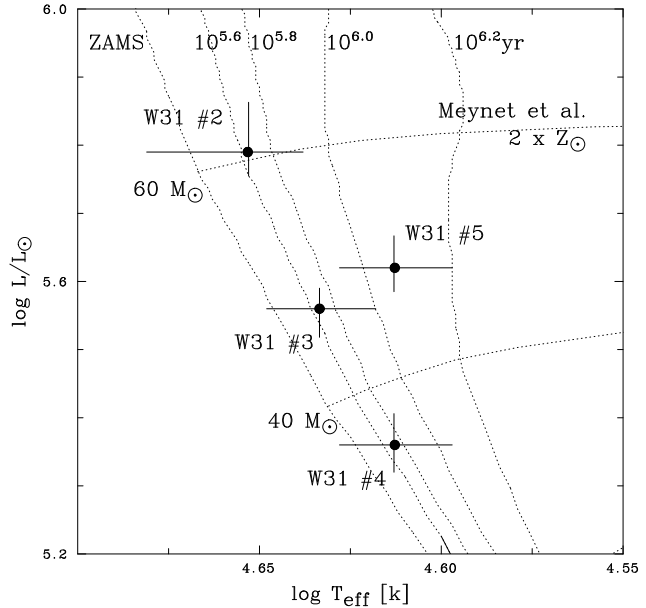}
\caption{Hertzsprung-Russell diagram of the 4 naked O stars in W31. 
\citet{Lejeune01}
isochrones are shown for ages of 10$^{3}$, 10$^{5.6}$, 10$^{5.8}$, 10$^{6.0}$ and 10$^{6.2}$
yr, based on the 2$\times Z_{\odot}$ \citet{Meynet94}  evolutionary 
tracks, from which a cluster age of $\sim$0.6 Myr is obtained.}
  \label{isos}
\end{figure}

Having established a slightly refined cluster distance, we now estimate 
stellar temperatures (and in turn luminosities) from spectroscopic fits to 
the near-IR diagnostics with the non-LTE CMFGEN code of \cite{Hillier98}.
CMFGEN solves the radiative transfer equation in the co-moving frame, 
under the additional constraint of statistical equilibrium.  Since CMFGEN 
does not solve the momentum equation, a density or velocity structure is 
required.  For the supersonic part, the velocity is parameterized with a 
classical $\beta$-type law, with an exponent of $\beta$=1 adopted. This is 
connected to a hydrostatic density structure at depth, such that the 
velocity and velocity gradient match at the interface. The subsonic 
velocity structure is set by a corresponding fully line-blanketed 
plane-parallel TLUSTY model \citep[v.200, see][]{Lanz03}. The atomic 
model is similar to that adopted in \citet{Hillier03}, including ions 
from H, He, C, N, O, Ne, Si, S, Ar, Ca and Fe, with metal abundances 
increased by a factor of two relative to those of the Sun from
\citet{Asplund04} for O, Ne and Ar, or \citet{Cox00} otherwise.

\begin{figure}
\begin{center}
\includegraphics[width=0.69\columnwidth,angle=270,clip]{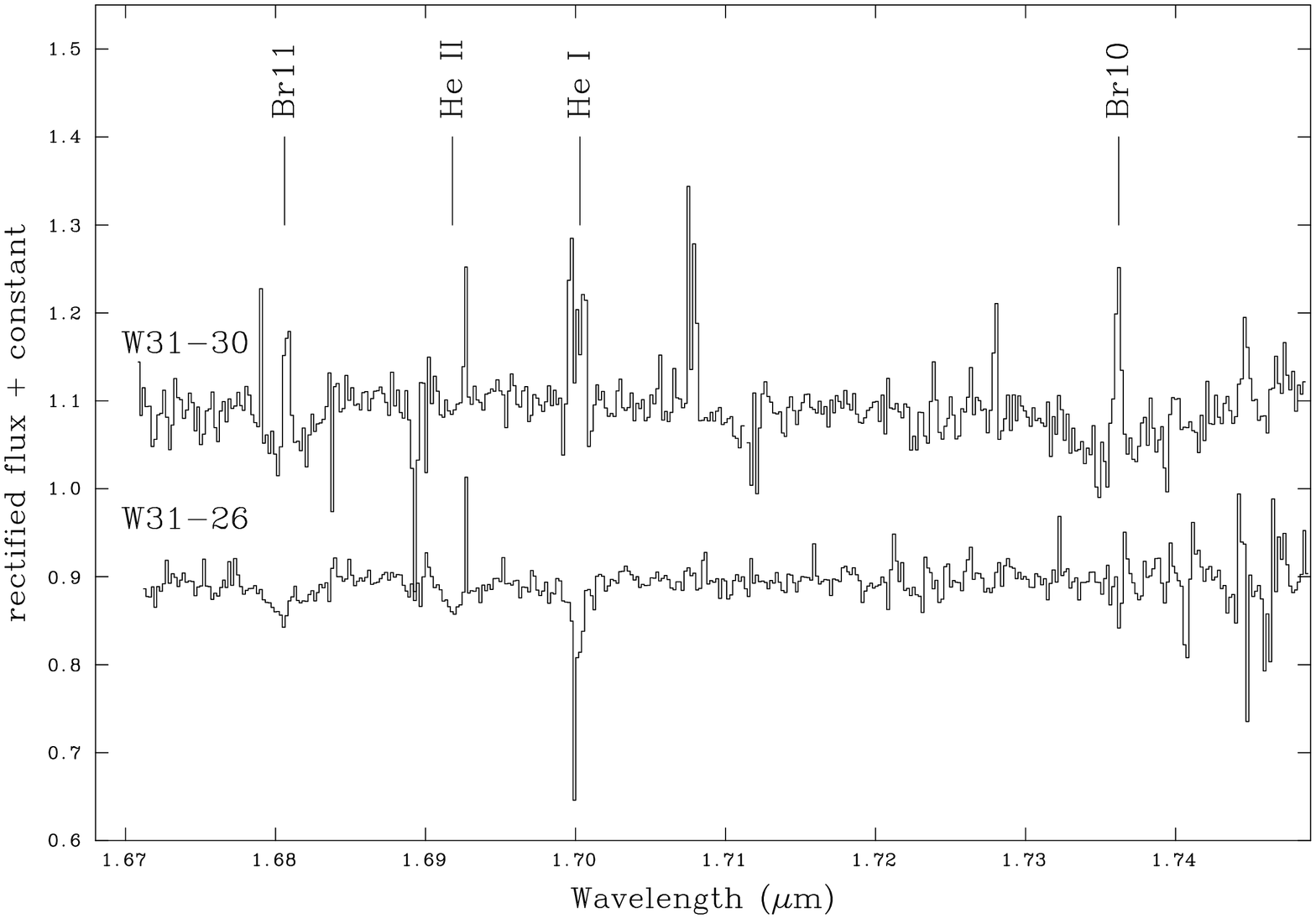}
\includegraphics[width=0.69\columnwidth,angle=270,clip]{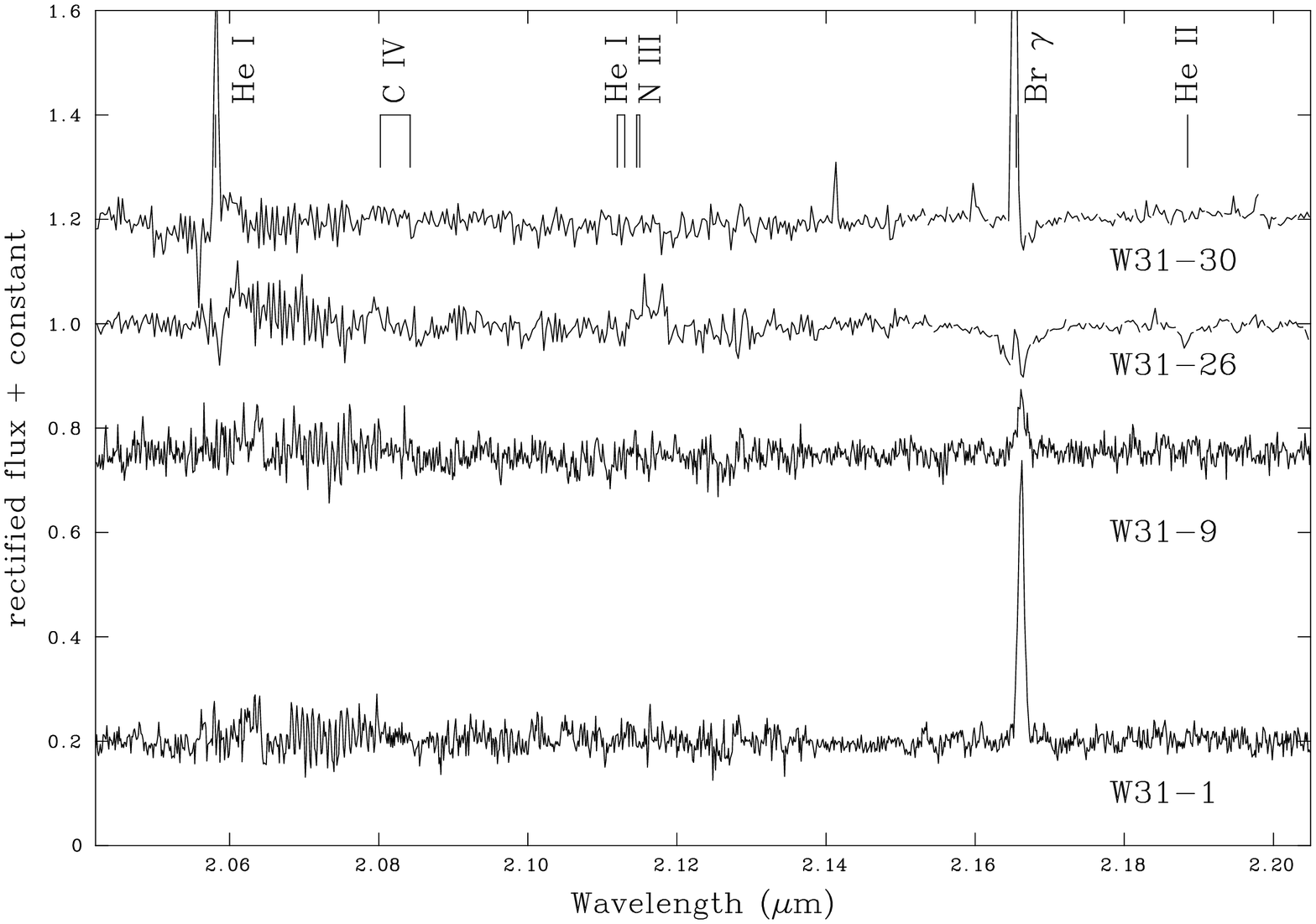} 
\caption{$H$ band (top panel) and $K$ band (lower panel) VLT/ISAAC spectra 
of the massive YSOs in W31. No $H$ band spectra are available for 
W31 \#1 and \#9. Br$\gamma$ is seen in emission throughout, while 
underlying photospheric absorption is also seen in \#26, demonstrating 
an evolutionary phase between the embedded and naked O star phases.}
\label{mysos}
\end{center}
\end{figure}

We have assumed a depth-independent Doppler profile for all lines when 
solving for the atmospheric structure in the co-moving frame, while in the 
final calculation of the emergent spectrum in the observer's frame, we 
have adopted a uniform turbulence of 50 km\,s$^{-1}$. Incoherent electron 
scattering and Stark broadening for hydrogen and helium lines are adopted. 
Finally, we convolve our synthetic spectrum with a rotational broadening 
profile. In view of the quality of our observations, rotational velocities
should be reliable to $\pm 50$ km\,s$^{-1}$.

For an adopted (uniform) terminal wind velocity of 2000 km\,s$^{-1}$, 
surface gravity of $\log g = 4$ and abundance ratio of He/H=0.1 by number, 
we varied the stellar radius and (non-clumped) mass-loss rate until an
acceptable match to the He\,{\sc ii} 1.692$\mu$m, He\,{\sc i} 1.700$\mu$m, 
Br$\gamma$ and He\,{\sc ii} 2.189$\mu$m was achieved. For the case of \#2
(O3--4\,V), for which negligible He\,{\sc i} 1.700$\mu$m absorption is detected,
we favour a temperature of $\sim$45\,kK since higher temperatures produce too weak
Br$\gamma$ absorption. Spectroscopic fits are presented in Fig.~\ref{fits}
while resulting physical and wind properties are shown in 
Table~\ref{stellar}.
Fits to diagnostic lines are generally satisfactory with the exception of
\#3 for which He\,{\sc ii} 2.189$\mu$m suffers from low S/N, such that
we rely upon the weaker 1.692$\mu$m line in this instance. Note that 
prominent nebular emission features are seen in the hydrogen Brackett series,
with the possible exception of \#2.

Our spectroscopic analysis did not yield a precise measurement of surface 
gravities (and in turn masses), so having established temperatures and 
luminosities, we overplot 2$\times Z_{\odot}$ theoretical isochrones from 
\citet{Lejeune01} in Figure~\ref{isos}, from which an age of $\sim$0.6 Myr 
was obtained, together with corresponding surface gravities and mass 
estimates (see Table~\ref{stellar}). Three of the four O stars lie along
a common isochrone of $\sim$0.5 Myr, although from Fig.~\ref{isos} \#5 
suggests a greater age of $\geq$1 Myr. However, it is possible that 
this source is a close double with equal mass components, whose absolute 
magnitudes would have $M_{\rm K}$ = --3.9 mag for a distance of
3.3 kpc. In this case, individual  components would have properties 
similar to those of \#4  in Table~\ref{stellar}, reducing the 
inferred cluster age to 
$\sim$0.5 Myr.

\subsection{Near-IR spectroscopy of massive YSOs}\label{mYSO}

In addition to the naked O stars, we have obtained ISAAC spectroscopy of 
four massive YSOs in W31 - high mass stars whose photospheric features are 
veiled by circumstellar dust with IR excesses due to hot dust. We observed 
\#1, \#9, \#26 and \#30 from \cite{Blum01} at both grating settings in the 
$K$-band spectra, but only \#26 and \#30 were observed in the $H$-band. These 
datasets are shown in Figure \ref{mysos}. In general, our observations 
confirm the lower quality datasets of \cite{Blum01}, with nebular Brackett 
emission lines, plus nebular He\,{\sc i} 1.700, 2.058$\mu$m for \#30.

\begin{figure*}
\includegraphics[width=0.69\textwidth,angle=270,clip]{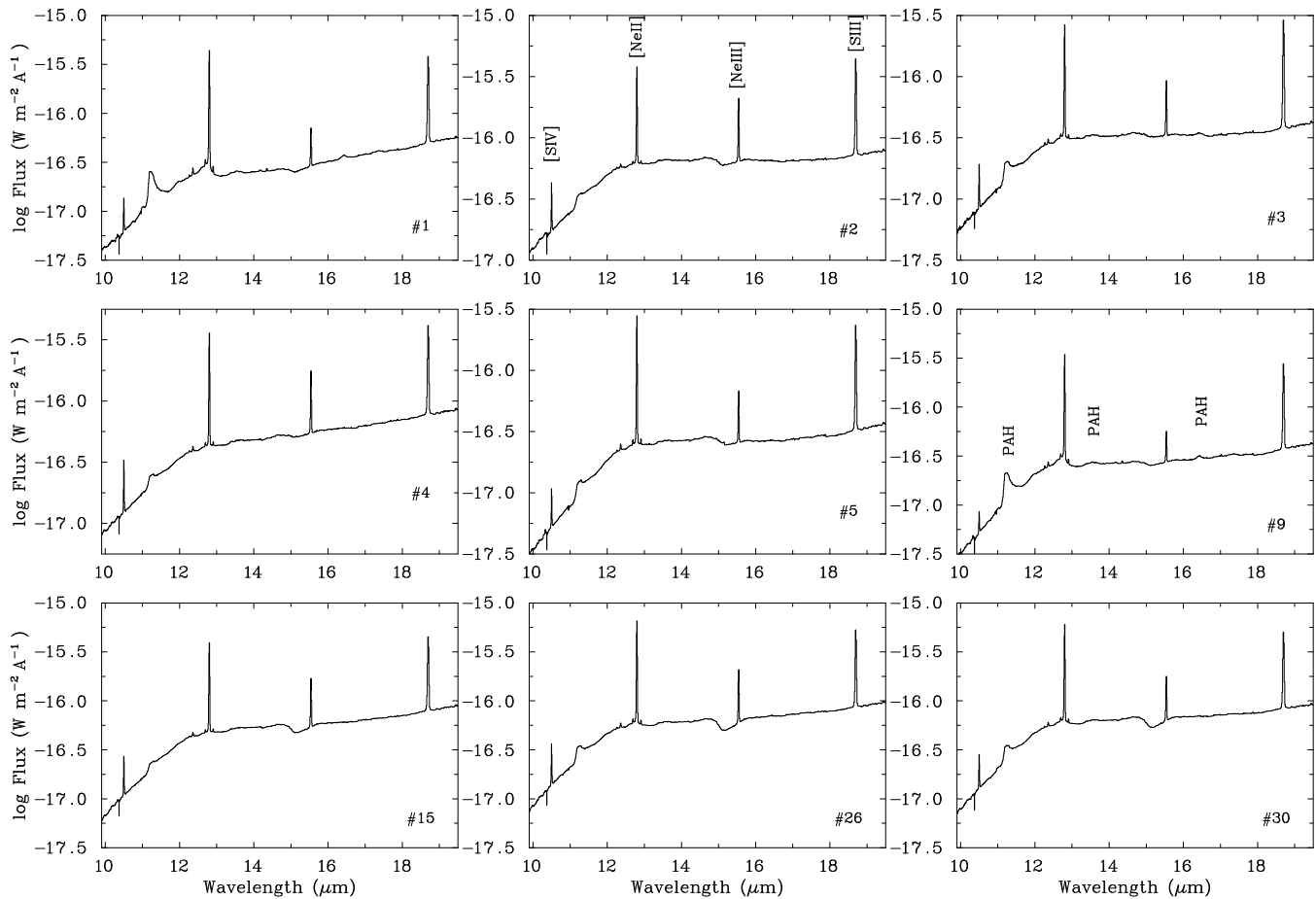} \caption{IRS 
short high (SH) medium resolution spectroscopy of W31 sources,
for which the fine structure lines of 
[S\,{\sc iv}], [Ne\,{\sc ii}], [Ne\,{\sc iii}] 
and [S\,{\sc iii}] are indicated in \#2, and the PAH (UIR) features 
are marked in \#9. These mid-IR datasets have been background 
subtracted using an OFF pointing.} 
\label{irs}
\end{figure*}

Uniquely among the massive YSOs, 
\#26 shows clear 
photospheric Br$\gamma$ absorption in addition to weak nebular Br$\gamma$ 
emission, plus He\,{\sc ii} 2.189$\mu$m, 
1.692$\mu$m and He\,{\sc i} 1.700$\mu$m absorption. 
From the observed
H-band classification diagnostics in the H-band we estimate an approximate
spectral type of O6, since $\log W_{\lambda}$ (He\,{\sc ii} 1.692)/$W_{\lambda}$ 
(He\,{\sc i} 1.700) $\sim$ --0.26. This star appears to be at the point of
revealing its photosphere, despite the presence of a strong IR dust
excess \citep[$H-K$ = 2.72 mag from][]{Blum01}.

In principle, we could estimate the physical properties of \#26, by
combining an estimate of its stellar temperature ($\sim$36\,kK) with the
$H$-band photometry from \cite{Blum01}, the mean $H$-band extinction for
the naked O stars ($A_{\rm H} \sim$ 3.4 mag) and the adopted distance
of 3.3 kpc. For this star the $H$-band is selected on the basis of 
significant dust emission in the $K$-band. Unfortunately, the resulting
parameters for  \#26 are unphysical since $\log L/L_{\odot} = 4.2$
would be obtained, i.e. mass of only 4$M_{\odot}$(!) for an assumed
surface gravity of $\log g$ = 4.0. More plausibly, \#26 may possess
a small dust excess, i.e. a large colour excess due to extreme extinction. 
Recall that this source lies close to the reddening line in the 
colour-magnitude diagram \citep{Blum01}. If the $(H-K)$ color 
were to arise solely from line-of-sight extinction, $A_{\rm H} \sim 7.9$ 
mag, from which $\log L/L_{\odot}$ = 6.2 would be obtained. In reality,
the physical properties of \#26 will lie between these two extreme cases.
Regardless, this star warrants  further study since it uniquely displays 
both stellar and circumstellar features amongst early-type stars in W31.

%\begin{landscape}
\begin{table*}
\begin{center}
\caption{Mid-IR fine structure line fluxes (intensities in bold font),
with units of 10$^{-14}$ W\,m$^{-2}$,
for the O stars and  massive YSOs from {\it Spitzer}/IRS spectroscopy of 
W31, 
together with archival {\it ISO}/SWS observations of selected \hii\ 
regions
from Peeters et al. (2002) 
%and the ONC from Simpson et al. (1998).
}
\nocite{Peeters02}
%\nocite{Simpson98}
\begin{tabular}{llllllccc}
\hline
             &  & \multicolumn{4}{c}{Line Fluxes ({\bf Intensities})} \\
Object       & A$_K$   & [SIV]  & [NeII]   & [NeIII]  &  [SIII]   & 
log$\frac{I[{\rm NeIII}]}{I[{\rm NeII}]}$ & log$\frac{I[{\rm 
SIV}]}{I[{\rm SIII}]}$ & log$\frac{I[{\rm SIV}]}{I[{\rm NeII}]}$ \\ 
             &         &  10.5\micron & 12.8\micron  &  15.5\micron    &   
18.7\micron \\
\hline
W31 \#1  &      &  \phantom{1}0.138$\pm$0.002 & \phantom{11}9.66$\pm$0.10  & \phantom{1}1.19$\pm$0.01 & 
11.4$\pm$0.1\\
& 2.1 &{\bf  \phantom{1}0.473$\pm$0.006}&{\bf \phantom{1}12.7$\pm$0.1}&{\bf 
\phantom{1}1.46$\pm$0.02}&{\bf 13.7$\pm$0.1}&{\bf -0.94$\pm$0.01}& 
{\bf --0.46$\pm$0.01} & {\bf --1.46$\pm$0.01} \\
W31 \#2  &      &  \phantom{1}0.452$\pm$0.004& \phantom{11}7.43$\pm$0.07  & \phantom{1}3.96$\pm$0.03 & 
12.7$\pm$0.1 \\
         & 2.1 
&{\bf \phantom{1}1.55$\pm$0.01}&{\bf \phantom{11}9.78$\pm$0.1}&{\bf 
\phantom{1}4.86$\pm$0.04}&{\bf 15.4$\pm$0.1}& 
{\bf --0.30$\pm$0.01} & {\bf -1.00$\pm$0.01} & {\bf -0.80$\pm$0.01} \\ 
W31 \#3  &      & \phantom{1}0.193$\pm$0.002&\phantom{11}5.44$\pm$0.05  & \phantom{1}1.61$\pm$0.02 &  
\phantom{1}8.7$\pm$0.1 \\
         & 2.1 & 
{\bf \phantom{1}0.66$\pm$0.01}&{\bf \phantom{11}7.2$\pm$0.1}&{\bf 
\phantom{1}2.0$\pm$0.02}&{\bf 10.5$\pm$0.1}& 
{\bf -0.56$\pm$0.01} & {\bf -1.20$\pm$0.01} & {\bf -1.03$\pm$0.01}\\  
W31 \#4  &      & \phantom{1}0.365$\pm$0.003&\phantom{11}7.34$\pm$0.07  & \phantom{1}3.14$\pm$0.04 & 
11.6$\pm$0.1 & \\
         & 2.1  
&{\bf \phantom{1}1.25$\pm$0.01}&{\bf \phantom{11}9.66$\pm$0.1}&{\bf 
\phantom{1}3.85$\pm$0.05}&{\bf 14.0$\pm$0.1}& 
{\bf -0.40$\pm$0.01} & {\bf -1.05$\pm$0.01} & {\bf -0.89$\pm$0.01}\\
W31 \#5  &      & \phantom{1}0.010$\pm$0.0001&\phantom{11}5.86$\pm$0.7  & \phantom{1}1.06$\pm$0.01 &  
\phantom{1}6.80$\pm$0.1 \\
         & 2.1  &{\bf \phantom{1}0.341$\pm$0.003}&{\bf \phantom{11}7.71$\pm$0.08}&{\bf 
\phantom{1}1.30$\pm$0.01}&{\bf \phantom{1}8.2$\pm$0.1}& {\bf -0.77$\pm$0.01} &{\bf -1.38$\pm$0.01} 
& 
{\bf -1.35$\pm$0.01} \\ 
W31 \#9  &    &  \phantom{1}0.062$\pm$0.001& \phantom{11}7.06$\pm$0.11  & \phantom{1}0.80$\pm$0.01 & 
\phantom{1}8.0$\pm$0.1 \\
         & 2.1  
&{\bf \phantom{1}0.211$\pm$0.003}&{\bf \phantom{11}9.3$\pm$0.1}&{\bf 
\phantom{1}0.99$\pm$0.01}&{\bf \phantom{1}9.7$\pm$0.1}& 
{\bf -0.97$\pm$0.01} & {\bf -1.66$\pm$0.01} & {\bf -1.64$\pm$0.01}\\
W31 \#15 &      & \phantom{1}0.301$\pm$0.002&\phantom{11}8.04$\pm$0.08  & \phantom{1}3.02$\pm$0.03 & 
12.8$\pm$0.1\\
         & 2.1  
&{\bf \phantom{1}1.03$\pm$0.01}&{\bf \phantom{1}10.6$\pm$0.1}&{\bf 
\phantom{1}3.71$\pm$0.04}&{\bf 15.4$\pm$0.1}&{\bf -0.46$\pm$0.01} &{\bf -1.18$\pm$0.01} 
&{\bf -1.01$\pm$0.01} \\
W31 \#26 &       &\phantom{1}0.396$\pm$0.004&\phantom{1}13.5$\pm$0.1 & \phantom{1}3.92$\pm$0.03 & 
15.2$\pm$0.2\\
         &2.1   &{\bf \phantom{1}1.36$\pm$0.01}&{\bf \phantom{1}17.8$\pm$0.2}&{\bf 
\phantom{1}4.81$\pm$0.04}&{\bf 18.3$\pm$0.2}& 
{\bf -0.57$\pm$0.01} &{\bf -1.13$\pm$0.01} &{\bf -1.12$\pm$0.01}\\ 
W31 \#30 &      &\phantom{1}0.286$\pm$0.002&\phantom{1}12.4$\pm$0.2  & \phantom{1}3.15$\pm$0.02 & 14.5$\pm$0.1\\
         & 2.1  
&{\bf \phantom{1}0.98$\pm$0.01}&{\bf \phantom{1}16.3$\pm$0.2}&{\bf 
\phantom{1}3.87$\pm$0.03}&{\bf 17.5$\pm$0.2}&
{\bf -0.63$\pm$0.01} & {\bf -1.25$\pm$0.01} & {\bf -1.22$\pm$0.01}\\ 
G23.96+0.15 &   &\phantom{1}0.33$\pm$0.08 & \phantom{1}34$\pm$1       & \phantom{1}2.4$\pm$0.2   &  
25.6$\pm$0.4 \\
           & 2.0&{\bf \phantom{1}1.3$\pm$0.2}&{\bf \phantom{1}46$\pm$1}&{\bf \phantom{1}3.0$\pm$0.3}&{\bf 
31.7$\pm$0.5} & {\bf -1.19$\pm$0.06} & 
{\bf -1.37$\pm$0.05} &{\bf -1.54$\pm$0.05} \\
G29.96$-$0.02 &  &\phantom{1}4.2$\pm$0.4 & \phantom{1}99$\pm$7       & 27 $\pm$1     &  46$\pm$1     \\
           &1.6  &{\bf 10.7$\pm$1} & {\bf 122$\pm$9} 
&{\bf 32$\pm$1}&{\bf 53$\pm$1}&{\bf -0.58$\pm$0.05 }
& {\bf -0.69$\pm$0.05} & {\bf -1.06$\pm$0.06} \\ 
% G110.10+0.05 & & 1.3$\pm
% 0.1$ & 14.7$\pm 0.5$ & 2.1$\pm 0.1$ & 16.3$\pm 0.4$ \\
% & 0.4 &    {\bf 1.6$\pm$0.1} & {\bf 15.5$\pm$0.5}& {\bf 2.2$\pm$0.1} & {\bf 
% 16.9$\pm$0.4} & {\bf --0.85$\pm$0.04} & {\bf --1.01$\pm$0.04} & {\bf 
% --0.97$\pm$0.04} \\
%ONC & 0.0 & \phantom{1}4.8$\pm 0.9^{a}$ & \phantom{1}12.9$\pm 1.5^{a}$ & 10.2$\pm 1.4^{a}$ & 
%14.6$\pm 1.6^{a}$ & --0.10$\pm$0.06 & --0.49$\pm$0.07 & --0.43$\pm$0.08 \\
\hline
\end{tabular} 
%\flushleft{$^1$ Fine structure lines taken from \cite{Peeters02}, 
%spectral type and A$_{\rm k}$ from \cite{Crowther08} \\ 
%$^2$ Fine structure lines taken from \cite{Peeters02}, spectral type from 
% \cite{Hanson227}, A$_{\rm k}$ from \cite{MartinH02a} }
\label{ratios}
\end{center}
%(a) units of $10^{-10}$ W\,cm$^{-2}$\,sr$^{-1}$
\end{table*}
%\end{landscape}

\section{Indirect properties of early-type stars in W31 from IRS 
spectroscopy}\label{finestruc}

\subsection{Mid-IR fine structure line ratios}

We present the {\it Spitzer} IRS short high medium resolution observations
of the naked O stars and massive YSOs in Fig~\ref{irs}. 
Although these datasets are background subtracted, no allowance for extended
nebular emission close to each source is made. Their continua are 
dominated by warm dust emission, while the solid state
Polycyclic Aromatic Hydrocarbon (PAH) features at  11.3$\mu$m, 13.6$\mu$m and 16.5$\mu$m --
previously known as Unidentified Infrared (UIR) emission -- are also seen (the 16.5$\mu$m
PAH feature is not detected in \#2). There is also a tentative detection of 15.2$\mu$m
absorption by icy CO$_2$ mantles on the silicate dust grains \citep{vanDishoek04}. Shallow
profiles with a soft blue wing starting near 14.9$\mu$m are reminiscent of the W33A massive
YSO \citep{Gibb04}, 
although we have no confirmation of the associated 4.3$\mu$m CO$_2$ absorption which is
strong in massive protostars.

\begin{figure}
\includegraphics[width=\columnwidth,clip]{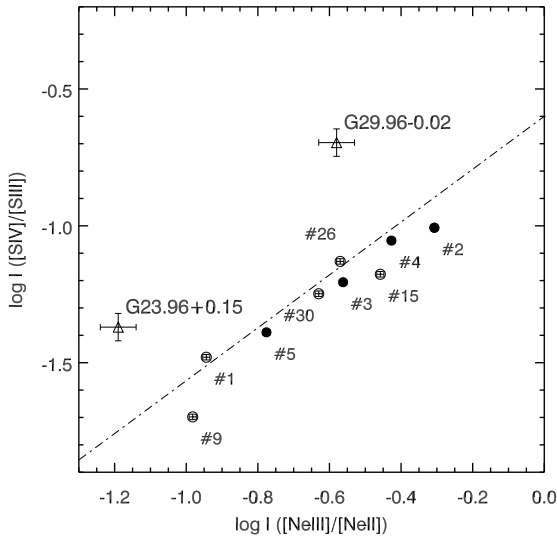}
\caption{Fine structure lines of the O-type stars (filled
circles) and massive YSOs (open circles) in W31,
with several compact and ultracompact H\,{\sc ii} regions 
discussed in this work included for comparison (open triangles). 
The dashed line is taken 
from \citet{MartinH02b}, and represents the best fit to 
observational datasets spanning a wide range of environments (see text).}
\label{emp_fstruc}
\end{figure}

We have measured the observed fluxes of the fine structure lines in all IRS
sources, using the ELF routine in DIPSO, which are presented in
Table \ref{ratios}. Other, weaker features are observed in the SH module, 
including Hu~$\alpha$ (H\,{\sc i} 7--6) at 12.37$\mu$m\footnote{At the 
spectral resolution of
IRS/SH the 12.37$\mu$m feature is actually a blend 
of H\,{\sc i} (7--6) and (11--8) for which
$I$(11--8)/$I$(7--6) = 0.123 according to Case B recombination theory
at $T_{e}$ = 7,500\,K and 10$^{4}$ cm$^{-3}$ \citep{Storey95}.}, 
for which $\log [I$(Hu $\alpha$)/$I$([Ne\,{\sc ii}] 12.8$\mu$m)]
= --1.92 $\pm$ 0.04. 
Where appropriate, we include de-reddened line intensities in Table~\ref{ratios}, 
obtained  from mean $K$-band extinctions measured previously together with the 
mid-IR extinction law of \cite{Morris}, namely 
\begin{eqnarray*}
A_{\rm 10.5\micron} & = & 0.637\,A_{K} \\
A_{\rm 12.8\micron} & = & 0.142\,A_{K} \\
A_{\rm 15.5\micron} & = & 0.106\,A_{K} \\
A_{\rm 18.7\micron} & = & 0.097\,A_{K}
\end{eqnarray*}
Of the mid-IR diagnostics, the [S\,{\sc iv}] line at 10.51 $\mu$m is 
significantly affected by the 9.7$\mu$m silicate feature. The 11.3$\mu$m 
PAH emission feature spans a range of equivalent
widths, typically $W_{\lambda}$ = 600$\pm$300 \AA, with the exception of two
massive YSOs, \#1 ($W_{\lambda} \approx$ 2500\AA) and \#9 ($W_{\lambda} 
\approx$
2100\AA). Weak UIR emission a 12.7$\mu$m can be seen in these two sources
(see Fig.~\ref{irs}).

In addition to the W31 sources, we also present mid-IR fine structure
line fluxes for
selected \hii\ regions in Table \ref{ratios}. Of these, G23.96+0.15 and
G29.96--0.02  lie within the inner Milky Way and possess both
mid-IR spectroscopy from Infrared Space Observatory \citep{Peeters02}
and near-IR spectroscopy of the ionizing star \citep{Hanson227, Crowther08}. 
Interstellar extinction for these were taken from \cite{Crowther08} and 
\cite{MartinH02a}, respectively. 
% We also include the 
% Orion Nebula Cluster (ONC) from \citet{Simpson98}.
% In view of the modest extinction towards the ONC plus instrumental 
% uncertainties presented by  \cite{Simpson98} we have not  applied any 
% mid-infrared interstellar extinction in this instance.

In Fig.~\ref{emp_fstruc} we compare the intensity ratios of
[Ne\,{\sc iii}]/[Ne\,{\sc ii}] and [S\,{\sc iv}]/[S\,{\sc iii}] 
for the W31 sources plus selected compact and \hii\ regions. The former
lie close to the best fit of observed ratios in H\,{\sc ii} regions in the 
Milky Way and Magellanic Clouds \citep{MartinH02b}, based on
observations presented by \citet{Peeters02} and \citet{Vermeij02}.
The compact and \hii\ regions are apparent outliers, but Fig.~1 from
\citet{MartinH02b} reveals a typical scatter of 0.2 dex and they 
did not correct line ratios for interstellar extinction. Physically,
these are probably amongst the highest density (ionization parameter)
H\,{\sc ii} regions of the full \citet{MartinH02b} sample.

The W31 sources with the highest ionization are the naked O stars \#2
(O3--4\,V) and \#4 (O5.5$\pm$0.5\,V) with massive YSOs \#1 and \#9 
possessing the lowest 
ionization nebulae, with a mixture of naked O stars and massive YSOs
at intermediate ionization. From these observations, \#15, \#26 and \#30
could be interpreted as possessing similar stellar properties to \#3 
(O5$^{+0.5}_{-1}$\,V) and \#5 (O5.5$\pm$0.5\,V) with later subtypes for 
\#1 and \#9, although we 
defer a formal estimate of their temperatures until \S~\ref{photoionization}. 
Despite this overlap, it may be significant that the naked O stars tend towards
higher ionization nebulae, i.e. higher temperatures (larger  stellar masses).

\begin{table*}
\begin{center}
\caption{Predicted mid-IR fine structure line ratios for inner Milky Way
H\,{\sc ii} regions. Elemental abundances are scaled to 1.5 $Z_{\odot}$
throughout. An  electron  density of 6,500\,cm$^{-3}$ is used for W31 on 
the basis of the radio-derived ionizing flux (10$^{50.4}$ ph\,s$^{-1}$) 
and radius of $\sim$30$''$ (0.5 pc at a distance of 3.3 kpc) for the 
giant H\,{\sc ii} region.}\label{table5}
\begin{tabular}{
l@{\hspace{2mm}}
c@{\hspace{2mm}}
c@{\hspace{2mm}}
c@{\hspace{2mm}}
c@{\hspace{2mm}}
r@{\hspace{2mm}}
c@{\hspace{2mm}}
c@{\hspace{2mm}}
r@{\hspace{2mm}}
r@{\hspace{2mm}}
r@{\hspace{2mm}}
r@{\hspace{2mm}}
r}
\hline
Star & Sp. Type & T$_{\rm eff}$ & $d$ & $\log Q_{0}$ & $n_{e}$ & 
$R_{s}$ & $\log U$ & log$\frac{I[{\rm NeIII}]}{I[{\rm NeII}]}$ &
log$\frac{I[{\rm SIV}]}{I[{\rm SIII}]}$ & log$\frac{I[{\rm ArIII}]}{I[{\rm NeII}]}$ &
log$\frac{I[{\rm SIV}]}{I[{\rm NeII}]}$ & log$\frac{I[{\rm Br}\alpha]}{I[{\rm NeII}]}$ \\
     &          & kK            & kpc &       & cm$^{-3}$ &  pc & \\
\hline
W31 \#2          & O3--4\,V      & 45  &3.3 & 49.6  & 6,500  & (0.24)&--1.6 &
1.01 & 0.42 & 0.52 & 1.22 & --0.01 \\
W31 \#3          & O5$^{+0.5}_{-1}$\,V      & 43  &3.3 & 49.3  & 6,500  & (0.19)&--1.7 &
0.65 & 0.23 & 0.28 & 0.81 & --0.28 \\
W31 \#4          & O5.5$\pm$0.5\,V    & 41  &3.3 & 49.0  & 6,500  & (0.15)&--1.8 &
0.24 & 0.01 & 0.05 & 0.39 & --0.52 \\
W31 \#5          & O5.5$\pm$0.5 \,V    & 41  &3.3 & 49.3  & 6,500  & (0.19)&--1.7 &
0.31 & 0.09 & 0.08 & 0.49 & --0.48 \\
G29.96--0.02     & O4--5\,V   & 41  &7.4 & 49.6  &20,000  & 0.13&--1.5  &
0.95 & 0.60 & 0.49 & 1.06 & 0.00 \\
G23.96+0.15      & O7.5\,V    & 38  &4.7 & 49.2  &70,000  & 0.05&--1.4 &
0.02 & 0.46 & --0.06 & 0.02 & --0.52 \\
\hline
\end{tabular}  
\label{logu}
\end{center}
\end{table*}

\subsection{Photoionization Modelling}\label{photoionization}

%\begin{figure}
%\includegraphics[width=\columnwidth,clip]{fstruc_w31_empi_cloudy.eps}
%\caption{As per Figure \ref{emp_fstruc}, with the CLOUDY model
%predictions as described in \S\ref{photoionization}.}
%  \label{cloudy_fstruc}
%\end{figure}

The primary aim of our study is to compare the direct near-IR stellar 
signatures of O stars in W31 (and \hii\ regions) with the indirect 
mid-IR nebular lines through predictions from photoionization models. 
We use version 08.00 of the photoionization code \textsc{cloudy},
last described by \cite{Cloudy}.
This solves the equations of thermal and statistical equilibrium for a 
model nebula,  represented by a sphere of gas with uniform density
$n$ and filling factor $\epsilon$ with a small central cavity which is
ionized and heated solely by the UV radiation of a single central star.

Nebular fluxes are predicted, given input abundances, ionizing flux 
distributions and physical parameters, most important of which is the
ionization parameter 
\begin{equation*}
U = Q_{0}/(4 \pi R^{2}_{\rm S} nc)
\end{equation*} 
and $Q_{0}$ is the number of  ionizing photons below
the H-Lyman edge at 912\AA. Here, $R_{\rm S}$ is the radius of the
Str\"{o}mgren sphere. Alternatively, 
\begin{equation*}
U = \frac{1}{c}\left(\frac{n \alpha_{\rm B}^2 Q_{0}}{36 
\pi}\right)^{1/3}
\end{equation*}
where $\alpha_{\rm B}$ is the Case B recombination coefficient.
For a given energy distribution of the ionizing 
radiation field, any 
combination of parameters which keeps $Q_{\rm H} n \epsilon^2$ constant 
will result in an identical ionization  structure of the gas 
\citep[see][]{Stasinska96}. 

\citet{Morisset04} have compared a number of stellar atmosphere codes to 
mid-IR observations of Milky Way H\,{\sc ii} regions, concluding that the 
non-LTE codes CMFGEN and WM-Basic \citep{Pauldrach01} provide the best 
match to observations \citep[see also][]{SimonDiaz08}. We therefore 
utilise  CMFGEN to provide the ionizing flux distributions, as discussed 
above.

Ideally, one would employ compact and ultra-compact H\,{\sc ii} 
regions that were both spherical and ionized by a single dominant source 
for such a study. However, such cases are incredibly rare, due to the 
lack of accurate subtypes for the ionizing stars and scarcity of 
space bourne mid-IR observations. G29.96--0.02 satisfies the ideal 
criteria relatively well, but G23.96+0.15 is irregular \citep{WC89},
and there is ionized gas throughout the W31 cluster. Fortunately, the O 
stars within W31 are relatively uniform in their ionizing output. In
contrast to the ONC, where $\theta^{1}$ Ori C dominates the extreme UV
radiation field, we assume that the ionized gas within IRS  apertures centred upon individual O 
stars are dominated by these stars. In reality, the diffuse radiation field
likely reflects a combination of the ionizing photons from multiple cluster
members, a considerations which should be bourne in mind in the following analysis.

\subsubsection{Nebular densities}

For the stars whose near-IR spectra enables a spectral type to be
determined, we can obtain empirical ionization parameters if the
distance and electron density is known. Table~\ref{logu} provides 
ionizing parameters for the two \hii\ regions whose ionizing stars have 
been determined from near-IR spectroscopy, G29.96--0.02 and G23.96+0.15,
implying $\log U \sim -1.5$ in both cases.

Specifically, for G29.96--0.02, \citet{Hanson227} estimated $\log Q_{0}$ = 
49.6 for a kinematic distance of 7.4 kpc, which together with a 
radio-derived Str\"{o}mgren radius of 3.5$''$ \citep[0.13 pc][]{WC89}  
requires a high  electron density of 20,000 cm$^{-3}$. This is typical of 
\hii\ regions, but is significantly higher than the density  of 817 
cm$^{-3}$ obtained from  ISO observations  using [O\,{\sc iii}] 
88/52$\mu$m by  \citet{MartinH02a}.  For G23.96+0.15, we adopt the 
kinematic distance of 4.7 kpc for which
\citet{Crowther08} obtained $\log Q_{0}$ = 49.4 and we
adopt a Str\"{o}mgren radius of 2$''$ (0.025 pc) from \citet{WC89},
although this \hii\ region has an irregular radio 
morphology. Again, the corresponding electron density of 70,000 cm$^{-3}$ 
is  much higher than the average density of 1543 cm$^{-3}$ obtained by  
\citet{MartinH02a}, arising in part due to the large aperture of ISO/LWS.

It was our intention that the IRS/LH spectroscopy of W31 
would have enabled us to deduce electron densities. Unfortunately,
due to severe saturation because of the bright dust continuum
this was not possible (\S~\ref{sect2.2}). 
We have therefore combined the radio-derived ionizing flux of $\log Q_{0}$
= 50.4 from \citet{Conti04} with the 30$''$ (0.5 pc) radius 
of 
the  H\,{\sc ii} region from diffuse Br$\gamma$ emission (recall 
Fig.~\ref{slits}) to estimate an electron density of 6,500\,cm$^{-3}$.
The ionization parameter implied in all cases is close to $\log U$ = 
--1.5, albeit slightly lower than for the \hii\ regions, with similar 
values anticipated for the massive YSOs in W31. For comparison,
\citet{Baldwin91} also derived $\log U = -1.5$ for the Orion Nebula
Cluster (ONC).

\begin{figure*}
\includegraphics[width=\columnwidth,clip]{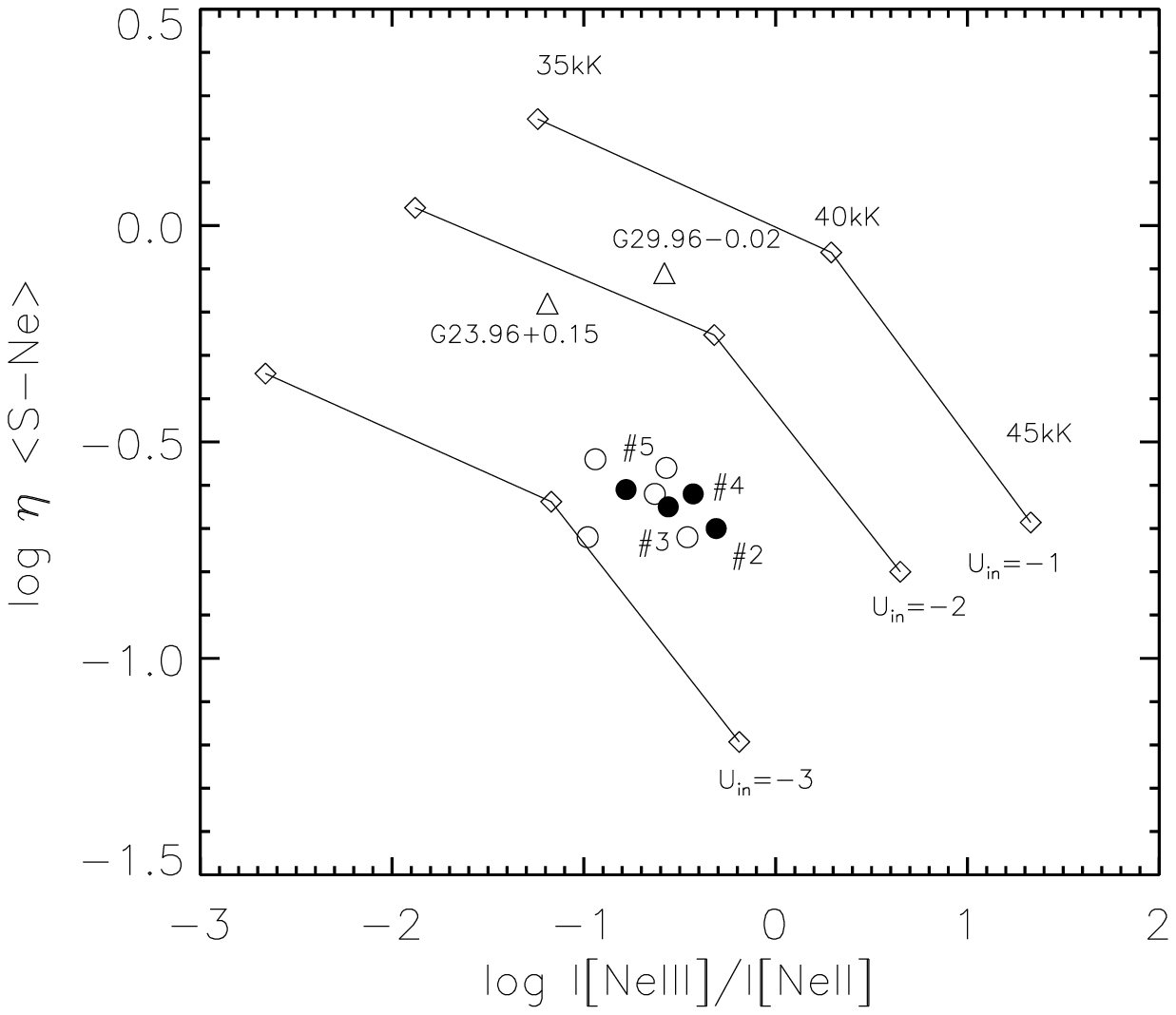}
\includegraphics[width=\columnwidth,clip]{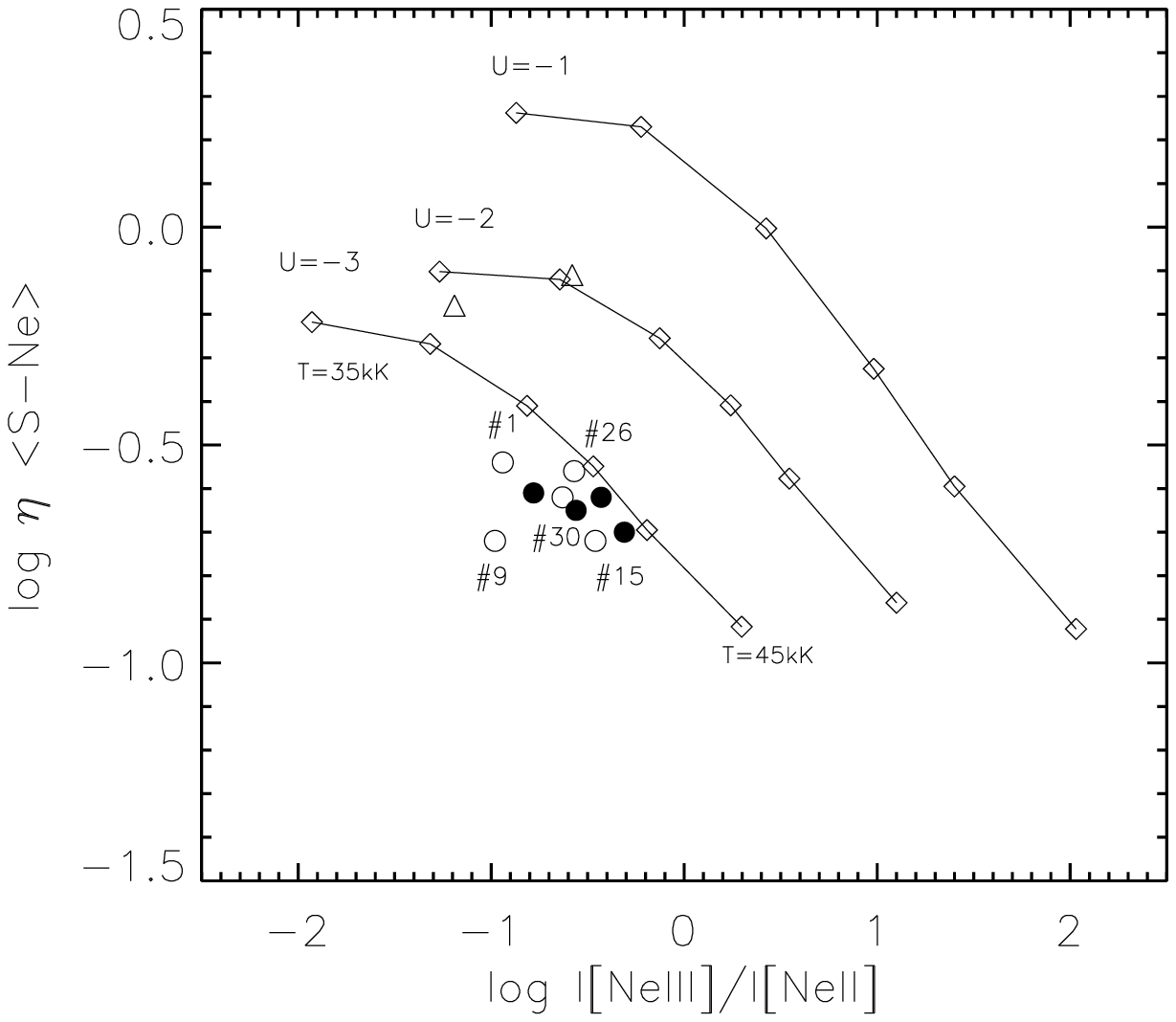} 
\caption{Log {\rm $\eta$}(S-Ne) versus log [NeIII]/[NeII] for solar 
metallicity models from \citet{SimonDiaz08} (left panel, $\log U_{\rm 
in}$ = --1, --2, --3) and this work (right panel, $\log U$ = --1, --2, 
--3), together with empirical data for naked O stars (filled
circles, identified in left panel) and massive YSOs (open circles,
identified in right panel) in 
W31, plus \hii\ regions 
(triangles, identified in left panel). See text for a discussion of 
the key differences between these photoionization models.} 
\label{ssdiaz_eta}
\end{figure*}

\subsubsection{Predicted mid-IR fine structure line ratios}

Table~\ref{logu} includes predicted mid-IR fine structure line ratios 
for each  case using \textsc{cloudy}. These were  obtained with
stellar models for individual stars from either \citet{Hanson227},
\citet{Crowther08} or the present study and were calculated with 
abundances scaled to 1.5 times the Solar value, with dust neglected.
From a comparison between the observed line ratios in Table~\ref{ratios}
and predicted values in Table~\ref{logu} it is apparent that agreement
is very poor. For example, the [Ne\,{\sc iii}] 15.5$\mu$m/[Ne\,{\sc ii}]
12.8$\mu$m and [S\,{\sc iv}] 10.5$\mu$m/[S\,{\sc iii}] 18.7$\mu$m 
ratios differs by 1.5$\pm$0.5 dex in all cases. This suggests either
that the stellar or photoionization models are at fault, or the problem
arises from local background variations close to individual W31 sources.
The former would appear to be more plausible, since the latter explanation
would not reconcile differences between ISO observations of \hii\ regions
and predictions.

Regardless of the origin for the discrepancy, 
we have calculated a grid of photoionization models for three
ionization parameters, $\log U$ = --1, --2 and --3 using stellar
atmospheric models appropriate for solar metallicity O dwarfs, as
listed in Table~\ref{params}. These were obtained from the O star 
calibration from Conti et al. (2008)\nocite{Conti_mono} together with the 
\citet*{Vink01} mass-loss prescription. A uniform rotational broadening of
200 km\,s$^{-1}$ was applied to synthetic spectra.
Dust was not included, but has little effect upon the predicted
line intensities with Orion-like dust grain compositions for dust-to-gas
mass ratios of 0.1--2\%. In the unlikely case of Milky Way ISM dust grains 
with a dust-to-gas mass ratio of 2\%, line intensity ratios may be 
affected by up to factor of two.

Observed line intensities are intrinsically linked to both ionization 
parameter,  $U$, and stellar temperature, $T_{\rm eff}$. One means of 
isolating the 
form of the radiation field is through the  ionization softness parameter
$\eta$ = ([O\,{\sc ii}]/[O\,{\sc iii}])/([S\,{\sc ii}]/[S\,{\sc iii}])
defined for H\,{\sc ii} regions detected optically by \cite{Vilchez88}. 
Analogously, the hardness of mid-IR nebular diagnostics can be
characterised by 
\begin{eqnarray*}
\eta({\rm S-Ne}) & = & \frac{([{\rm S\,IV}] 10.5\mu\,m/[{\rm S\,III}] 
18.7\mu\,m)}{([{\rm Ne\,III}] 15.5\mu\,m/[{\rm Ne\,II}] 12.8\mu\,m) }
\end{eqnarray*}
as discussed by Morisset (2004). Variations in $U$ and $T_{\rm eff}$ are 
broadly independent in a plot of  $\eta$(S--Ne) versus [Ne\,{\sc 
iii}]/[Ne\,{\sc ii}]. In  Fig.~\ref{ssdiaz_eta} we compare our 
\textsc{cloudy} predictions for  Solar  composition nebulae for $\log U$ = 
--1, --2, --3  to the observed positions of W31 stars and \hii\ 
regions, together with recent predictions  from \cite{SimonDiaz08} which 
are also based on \textsc{cloudy} and CMFGEN models. The W31 sources
lie very close together in Fig.~\ref{ssdiaz_eta}, with the marginal
exception of the massive YSO \#9. This source is also an outlier in 
Fig.~\ref{emp_fstruc}, possibly as a result of a reduced density
(ionization parameter).

%\begin{figure}
%\includegraphics[height=\columnwidth,angle=-90,clip]{fnu.eps}
%\caption{Synthetic UV spectra from $2 Z_{\odot}$ stellar models
%for O3, O5, O7 and O9V subtypes (from top to bottom) in which
%key ionization thresholds are indicated.}
%\label{fnu}
%\end{figure}

G29.96--0.02 lies close to the $(T_{\rm eff}, \log U_{\rm in})$ = (40kK, 
--2)  photoionization model from  \cite{SimonDiaz08}, in reasonable 
agreement with the 
empirical properties of (41kK, --1.5) from Table~\ref{logu}. In contrast, 
our results predict 
rather  poor agreement for G29.96--0.02 since it sits close to the (37kK, 
--2) photoionization model. The reason for this difference is that 
\cite{SimonDiaz08} adopt a non-standard model dependent definition of 
ionization parameter, namely
\begin{equation*}
U_{\rm in} = Q_{0}/(4 \pi R^{2}_{\rm in} nc)
\end{equation*} 
which is based upon the {\it inner} radius of the model cloud, $R_{\rm 
in}$ for  the photoionization calculation rather than the Str\"omgren 
radius, $R_{\rm S}$.
Consequently, the poor agreement for G29.96--0.02 obtained here simply 
confirms the previous (poor) consistency discussed by \citet{Morisset02}
and \citet{MartinH02b}.
The comparison for G23.96+0.15 fares little better, with ($T_{\rm eff}, 
\log 
U)$ = (36kK, --2) predicted, in contrast to empirical estimates of (38kK, 
--1.4) from Table~\ref{logu}. The situation for the W31 O stars is fairly 
similar, with
ionization parameters of $\log U < $--3 predicted in the
right panel of Fig.~\ref{ssdiaz_eta}, yet  --1.7 is more typical of
compact and \hii\ regions (Table~\ref{table5}).

In \S~\ref{distances}, we indicated that W31 is anticipated 
to be metal-rich with respect to the Solar neighbourhood. Similar 
arguments apply for G23.96+0.15 and especially  G29.96--0.02. For 
a kinematic distance of  $\sim$7 kpc to G29.96--0.02, this \hii\ region 
would lie at $R_{\rm  GC}$ =  4 kpc from the  Galactic 
Centre\footnote{\citet{Sewilo04} obtained a distance
of 8.9 kpc or $R_{\rm GC}$ = 4.5 kpc to G29.96--0.02.}, versus  3.7 kpc 
for G23.96+0.15 for a kinematic distance of 4.7 kpc.

% 5.7 or 8.2 for 8kpc 220 km/s using 97.4 V_LSR (1996A&AS 115 81)
% sewilo 8.9 using 91.3 V_LSR 

Therefore, we have also calculated a set 
of super-solar (2 $Z_{\odot}$) \textsc{cloudy} models, based upon 
O star models that are identical to the solar composition grid
except that both metal abundances and mass-loss rates are increased
(see Table~\ref{params} for the latter). 
%Synthetic UV spectra for
%O3--09V subtypes are presented in Fig.~\ref{fnu} in which
%ionization thresholds that are key to this study are shown. 
Solely early O stars provide significant numbers of extreme UV photons 
capable of producing [Ne\,{\sc iii}] and [S\,{\sc iv}] nebular emission
while [Ne\,{\sc ii}] and [S\,{\sc iii}] emission is expected to cease
beyond $\sim$O9.5V.

\begin{figure}
\includegraphics[width=\columnwidth,clip]{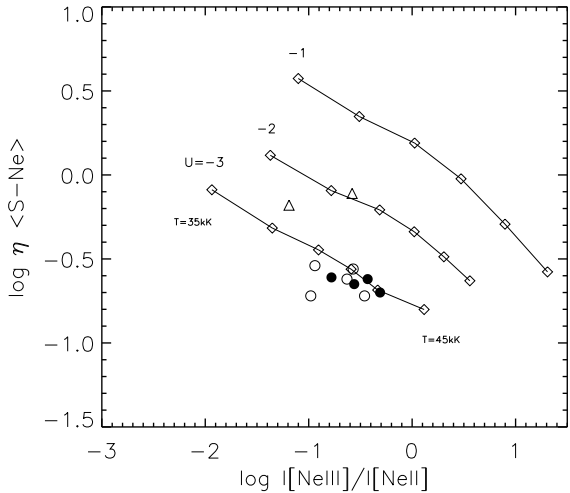}
\caption{log $\eta$(S-Ne) versus log I[NeIII]/I[NeII] ratio for 2$\times$ 
Solar metallicity grid for $\log U = -1, -2, -3$. Symbols have
the same meaning as in Fig.~8.}
  \label{eta_2z}
\end{figure}

Predictions from $2 Z_{\odot}$ photoionization models are presented in 
Fig.~\ref{eta_2z}.  
In general, from a comparison with the Solar models, these yield 
slightly higher stellar  temperatures and ionization parameters. G29.96--0.02
lies close to $(T_{\rm eff}, \log U)$ = (38kK, --2), representing a
slight improvement with respect to the empirical temperature. Similar
comments apply for  G23.96+0.15, although predicted temperatures remain 
too low, and ionization parameters are offset by --1.0 to --1.5 dex for 
G23.96+0.15 and the O stars in W31.

\begin{table}
\begin{center}
\caption{Stellar parameters for solar (twice solar) O dwarf model 
atmospheres used in photoionization calculations, based upon calibrations
presented in Conti et al. (2008) with mass-loss rates obtained from the 
Vink et al. (2001) prescriptions.}\label{params}
\begin{tabular}{l
@{\hspace{2mm}}
l@{\hspace{2mm}}
l@{\hspace{2mm}}
l@{\hspace{2mm}}
l@{\hspace{2mm}}
l@{\hspace{2mm}}
l@{\hspace{2mm}}l}
\hline
Subtype & T$_{\rm eff}$ & $\log L$   & $M$         & $\log Q_{0}$ & 
$v_{\infty}$ & \multicolumn{2}{c}{$\log \dot{M}$} \\
        & kK            & $L_{\odot}$& $M_{\odot}$ &    & km\,s$^{-1}$
& \multicolumn{2}{c}{$M_{\odot}$\,yr$^{-1}$}\\
\hline
O3 V & 45 & 5.88 & 74 & 49.65 & 3200 & --5.41 (--5.15) \\
O4 V & 43 & 5.77 & 64 & 49.5  & 3000 & --5.55 (--5.30) \\
O5 V & 41 & 5.57 & 51 & 49.25 & 2900 & --5.86 (--5.61) \\
O6 V & 39 & 5.39 & 41 & 49.0  & 2600 & --6.12 (--5.86) \\
O7 V & 37 & 5.25 & 36 & 48.8  & 2300 & --6.33 (--6.07) \\
O8 V & 35 & 5.10 & 31 & 48.55 & 1750 & --6.49 (--6.23) \\
O9 V & 33 & 4.91 & 25 & 48.2  & 1500 & --6.79 (--6.53) \\
O9.5 V&31.5& 4.78 & 23 & 48.0 & 1200 &--6.97 (--6.72) \\
\hline
\end{tabular}  
\end{center}
\end{table}

% W31-2 41 -> 45: 4
% W31-3 39 -> 43: 4
% W31-4 40 -> 41: 1
% W31-5 39 -> 41: 2
% G23.96 36 -> 38: 2

\begin{figure}
\includegraphics[width=\columnwidth,clip]{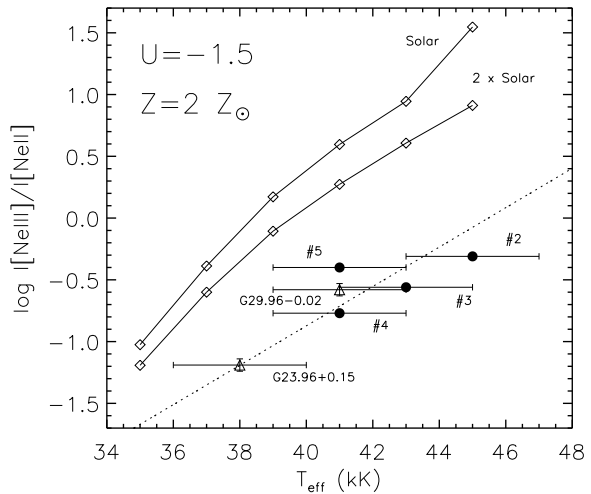}
\caption{Comparison between the predicted [Ne\,{\sc iii}]/[Ne\,{\sc 
ii}] intensity ratio and effective temperature
for $\log U$=-1.5 models at Solar and 
2$\times$Solar composition.
Empirical data are naked O stars in W31 (filled circles) and inner
Milky Way \hii\ regions (open triangles) together with a first order
fit to the empirical datasets (dotted line, see text).
% (lower panel) 
% As above, for outer Milky Way sources (star for ONC, 
% diamond for G110.10+0.05) and a +1.5\,kK offset to the Solar metallicity 
% grid (dotted line).
} 
\label{Neon}
\end{figure}

\begin{table*}
\begin{center}
\caption{Estimates of spectral types of the ionizing stars 
responsible for 
compact and ultra-compact H\,{\sc ii} regions 
in the inner Milky Way ($R_{\rm GC} < 7$ kpc) from
Peeters et al. (2002) plus massive YSOs in W31 from the present study. 
Interstellar extinctions are taken from the present study,
\citet{MartinH02a}, or $A_{\rm K}$ = 1.6 mag otherwise (marked with $\ast$).
Kinematic distances  adopt a  Solar galactocentric distance of 8.0 kpc \citep{Reid93}. 
Temperatures and spectral types are  obtained for both the  [Ne\,{\sc  
iii}]/[Ne\,{\sc ii}] and [S\,{\sc iv}]/[Ne\,{\sc ii}] diagnostic ratios.} 
\begin{tabular}{
l@{\hspace{2mm}}
l@{\hspace{2mm}}
l@{\hspace{2mm}}
c@{\hspace{2mm}}
c@{\hspace{2mm}}
c@{\hspace{2mm}}
c@{\hspace{2mm}}
c@{\hspace{2mm}}
c@{\hspace{2mm}}
c@{\hspace{2mm}}
c@{\hspace{2mm}}
c@{\hspace{2mm}}
c@{\hspace{2mm}}
c}
\hline
Source & IRAS & Alias & $V_{\rm LSR}$ & $d$ & $R_{\rm GC}$ & Ref& $A_{K}$ & $\log$  & 
$T_{\rm eff}$ & Sp Type & $\log$ & $T_{\rm eff}$ & Sp Type\\
      &&& km\,s$^{-1}$ & kpc & kpc    &      & mag  &  $\frac{[Ne\,III]}{[Ne\,II]}$
  & kK & & $\frac{[S\,IV]}{[Ne\,II]}$ & kK \\
\hline
%\multicolumn{14}{c}{$R_{\rm GC} < 7$ kpc} \\
G1.13--0.11 & 17455--2800 & Sgr D& --17.1 & \phantom{0}8.0 & \phantom{0}0.2 & a   & 
1.8\phantom{$\ast$} & --0.36 & 43 & O4 V 
& --0.69 & 43 & O4 V\\
W31 \#1 & & & & \phantom{0}3.3 & \phantom{0}4.8 & k & 2.1 & --0.94 & 39.5 & 
O6 V & --1.46 & 39 & O6 V \\
W31 \#9 & & & & \phantom{0}3.3 &  \phantom{0}4.8& k & 2.1 & --0.97 & 39.5 & 
O6 V & --1.64 & 38 & O6.5 V \\
W31 \#15 & & & & \phantom{0}3.3 &\phantom{0}4.8 & k & 2.1 & --0.46 & 42.5 & O4 V & --1.01 & 
41.5 & O5 V \\
W31 \#26 & & & & \phantom{0}3.3 &\phantom{0}4.8 & k & 2.1 & --0.57 & 42 & 
O4--5 V & --1.12 & 41 & O5 V \\
W31 \#30 & & & & \phantom{0}3.3 &\phantom{0}4.8 & k & 2.1 & --0.63 & 41.5 & O5 V & --1.22 & 
40 & O5.5 V \\
G33.91+0.11 & 18502+0051  &       & --100.0 & \phantom{0}6.6 &  \phantom{0}4.5  & b, c& 
1.6$^{\ast}$ & --0.66 & 42 & O4--5 V & 
--1.18 & 41  & O5 V  \\
G49.20--0.35 & 19207+1410 &       & --67.9 & \phantom{0}5.2 & \phantom{0}6.1 & d & 
1.6$^{\ast}$ & +0.03 & 45 & O3 V & --0.23 
& 45  & O3 V \\
G301.11+0.97 &12331--6134 &   & --40.8 & \phantom{0}4.1 & \phantom{0}6.8 & h & 1.6$^{\ast}$ 
& --0.16 & 44 &O3--4 V & --0.38 &  44  
& O3--4 V\\
G326.44+0.91 &15384--5348 &   & --41.0 & \phantom{0}2.5 & \phantom{0}6.1 & h & 
1.3\phantom{$\ast$}   & --0.55 & 42 &O4--5 V & 
--0.86  & 42  & O4--5 V \\
G328.31+0.43 &15502--5302 &   & --91.7 & \phantom{0}5.4 & \phantom{0}4.4 & h & 
2.7\phantom{$\ast$}   & --0.63 & 42 &O4--5 V & 
--1.18 & 41   & O5 V  \\
G332.15--0.45 &16128--5109 &   & --55.7 & \phantom{0}3.5 & \phantom{0}5.2 & h & 1.6$^{\ast}$ 
& --0.26 & 43.5&O4 V & --0.66  & 
43 & O4 V 
\\
G351.46--0.44 &17221--3619&    &--22.3 & \phantom{0}3.3 & \phantom{0}4.8 & h & 
1.8\phantom{$\ast$} & --1.41 & 39 & O6 V 
& --1.54 & 40 & O5.5 V  \\
%\multicolumn{14}{c}{$R_{\rm GC} \geq 7$ kpc} \\
%G70.29+1.60 & 19598+3324 & K3--50A& --24.2 & \phantom{0}8.1 & \phantom{0}9.3 & e & 1.6\phantom{$\ast$}      & +0.36 & 41.5  & 
%O5 V  & --0.23 & 40  & O5--6 V  \\
%G93.53+1.47 &21190+5140 &M1--78 &--73.8 & \phantom{0}9.0 & 12.4 & f & 0.0\phantom{$\ast$}      & +0.00 & 39  &O6V & --0.68 &   
%&\\
% G110.10+0.05 &23030+5958 & IC 1470 &--49.5 & \phantom{0}5.0 & 10.8 & g & 
% 0.4\phantom{$\ast$}  & --0.85 & 37 & O7V  & --0.97 & 
% 37.5 & O7V 
%\\
%G111.62+0.37 &23133+6050 & &--56.3 & \phantom{0}5.6 & 11.3 & h & 0.0\phantom{$\ast$}      & --1.63 & 34.5  & O8V  & --1.80 & 
%35  & O8V  \\
%G133.70+1.20 &02219+6152 & W3A & --36.5 & \phantom{0}3.4 & 10.6 & i & 1.5\phantom{$\ast$}      & +0.37 & 42   & O4--5V  & 
%+0.15   & 41.5   & O5V 
%\\
%G289.88--0.79 &10589--6034 &   &+20.2 & \phantom{0}7.7 & \phantom{0}9.0 & h & 1.5\phantom{$\ast$}      & --0.55 & 38 & O6.5V   
%& --0.67  & 39   & O6V\\
%G291.86--0.68 &11143--6113 &   &+25.0 & \phantom{0}8.6 & \phantom{0}9.3 & j & 1.6$^{\ast}$ & +0.64 &43 & O4V  &  +0.84 & 44  & 
%O3--4V\\
%G298.19--0.78 &12063--6259 &   &+22.9 & \phantom{0}9.8 & \phantom{0}9.3 & h & 0.8\phantom{$\ast$}    & +0.30 & 41  & O5V  & 
%+0.15 & 41  & O5V  \\
%G298.23--0.33 &12073--6233 &   &+31.0 &10.5 & \phantom{0}9.7 & j & 0.8\phantom{$\ast$}    & +0.80 & 43& O4V  & +0.78 & 44  &  
%O3--4V \\
\hline
\end{tabular}  
\label{ne3_ne2}
\end{center}
(a) \cite{Mehringer98}; (b) \cite{WC89}; (c) \cite{Fish03}; (d) \cite{Sewilo04}; (e) \cite{Araya}; (f) 
\cite{MD92}; (g) \cite{Quireza05}; (h) \cite*{Bronfman96}; (i) 
\cite{Afflerbach96}; (j) \cite{Caswell87}; (k) this study
\end{table*}

\subsection{[Ne\,{\sc iii}]/[Ne\,{\sc ii}] calibrations} \label{emp}

Let us now focus upon the predicted  [Ne\,{\sc iii}]/[Ne\,{\sc ii}] ratio 
versus stellar temperature, which is presented in Fig~\ref{Neon} for the 
Solar and  2$\times Z_{\odot}$ metallicity grids at $\log U = -1.5$. 
We have selected this value for the ionization parameter since this is
characteristic of compact and especially \hii\ regions in our sample.
Empirical  results are also shown, with stellar temperatures drawn
from Table~\ref{table5}, quantifying earlier discrepancies
with respect to the predictions. An explanation for
the disagreement is not readily apparent, 
although it may involve incomplete line
blanketing in current non-plane parallel model atmospheres at high 
energies. Nevertheless, observations allow us to estimate the
temperatures of the ionizing O stars in other dense
H\,{\sc ii} regions {\bf (for which $\log U \sim -1.5$)} within the inner Milky Way,
to within $\pm$2\,kK (i.e. one spectral subtype) in principle. A first order  
fit to the empirical datasets using {\sc IDL's} polyfit routine reveals
\begin{eqnarray*}
(T_{\rm eff}/kK)_{2 Z_{\odot}} & = & +45.47 \pm 1.37 \\
                               &   & +6.27 \pm 1.96
(\log I[{\rm Ne\,III}]/I[{\rm Ne\,II}]).
\end{eqnarray*}
O subtypes may then be estimated from the recent $T_{\rm eff}$ calibration 
of \citet*{Martins05}. Of course, this approach is only practical for compact
and \hii\ regions ionized by early- and mid- O 
stars, in view of the weakness of [Ne\,{\sc iii}] 15.5$\mu$m for late O-types.

%(T_{\rm eff}/kK)_{2 Z_{\odot}} & = & +44.74 \pm 0.13 \\
%                               &   & +4.93 \pm 0.17
%(\log I[{\rm Ne\,III}]/I[{\rm Ne\,II}]) \\
%                               &   & +0.69 \pm 0.10
%(\log I[{\rm Ne\,III}]/I[{\rm Ne\,II}])^2.

To illustrate its potential diagnostic role, in Table~\ref{ne3_ne2} we 
provide stellar temperatures derived from extinction corrected [Ne\,{\sc 
iii}]/[Ne\,{\sc ii}] ratios for compact and ultra-compact H\,{\sc ii} 
regions from ISO/SWS observations \citep{Peeters02}, plus the massive 
YSO's in W31 from the present study. Extinctions are either taken from 
\cite{MartinH02a}, or $A_{K}$ = 1.6 mag is adopted otherwise. We omit the 
peculiar compact H\,{\sc ii} region G93.53+1.47 (M1--78) from this sample 
\citep{MartinH08}. For the massive YSO's in W31, Table~\ref{ne3_ne2} 
presents spectral types inferred from the mean interstellar extinction 
obtained from the naked W31 O stars. In reality, higher extinctions may be 
expected for these cases (recall \S~\ref{mYSO}). Fortunately, the use 
of a higher extinction does not affect the resulting stellar 
temperatures/subtypes. For \#26, a subtype of O4--5V is obtained, versus an 
approximate subtype of O6V from its H-band ISAAC spectrum 
(\S~\ref{mYSO}).

Since our calibration has been established using solely compact H\,{\sc 
ii} regions within the inner Milky Way - with super-solar elemental
abundances -- one should consider separately the case of H\,{\sc ii} regions close to, 
or exterior to, the Solar circle. Of course, metal content is 
very relevant to gas cooling and metal-line blanketing of the extreme UV 
energy distributions of O stars. Unfortunately, among the compact H\,{\sc ii} regions 
close to the Solar circle, very few possess both mid-IR spectroscopy and
well determined properties for the dominant ionizing star. The ONC --  for which $\theta^{1}$ Ori C (HD 37022,  O6-7\,Vp) is the 
dominant source of ionizing radiation -- is one such case in that it possesses both mid-infrared spectroscopy
\citet{Simpson98} and a contemporary analysis of its stellar content by 
\citet{SimonDiaz06}.

Other examples are generally complicated by the presence of multiple ionizing
stars of uncertain spectral type (e.g. NGC~7538) or mid-IR spectroscopy is lacking.
The only other case for which mid-IR spectroscopy is available together with 
a well determined spectral type of the ionizing star is  G110.10+0.05  (IC~1470, Sh~2--156)
for which \citet{Hunter90} derived O6.5\,V for its ionizing star. Therefore, 
we refrain from attempting a solar metallicity calibration at this time. However, 
we are able to compare the ONC with the \textsc{cloudy} predictions, recalling 
that \citet{Baldwin91} derived a value of $\log U$ = --1.48 for the ONC. 
In contrast to the metal-rich H,{\sc ii} regions, the observed neon ratio for 
the ONC {\it is} in better agreement with the solar-metallicity predictions for 
$T_{\rm eff}$ = 39$\pm$1\,kK for $\theta^{1}$ Ori C \citep{SimonDiaz06}. 

\subsection{[S\,{\sc iv}]/[Ne\,{\sc ii}] and [Ar\,{\sc iii}]/[Ne\,{\sc ii}] calibrations} \label{emp}

\begin{figure}
\includegraphics[width=\columnwidth, clip]{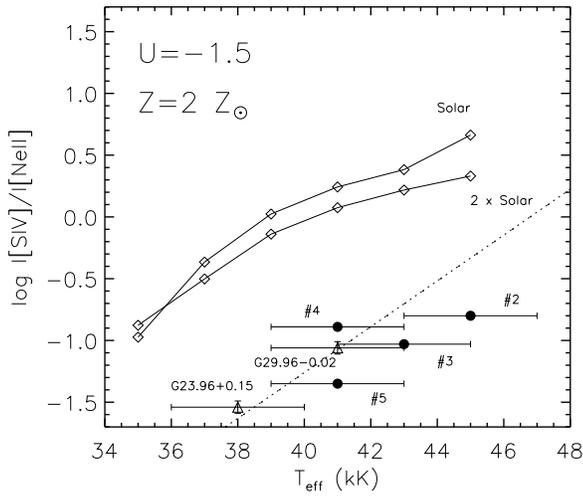} 
\caption{Effective temperature versus log 
[S\,{\sc iv}]/[Ne\,{\sc ii}] for our $\log U=-1.5$ solar and 2$\times$ 
Solar models
together with inner Milky Way sources and resulting empirical calibration (dotted-line).
%(Lower panel) As above, for outer Milky Way sources and our empirical
%calibration (dotted-line)
}
\label{U_plot}
\end{figure}

Of course, neither $\eta$(S--Ne) nor [Ne\,{\sc iii}]/[Ne\,{\sc ii}]
are available from ground-based observations. In such cases, only
[Ar\,{\sc iii}] 8.9$\mu$m, [S\,{\sc iv}] 10.5$\mu$m and [Ne\,{\sc ii}] 
12.8$\mu$m are accessible. In Figure \ref{U_plot} we compare the [S\,{\sc 
iv}]/[Ne\,{\sc ii}] ratio versus stellar temperature for sources in W31, 
G29.96--0.02 and  G23.96+0.15, plus \textsc{cloudy} predictions for 
2$\times$Solar stellar models. 

\begin{figure}
\begin{center}
\includegraphics[width=0.8\columnwidth,clip]{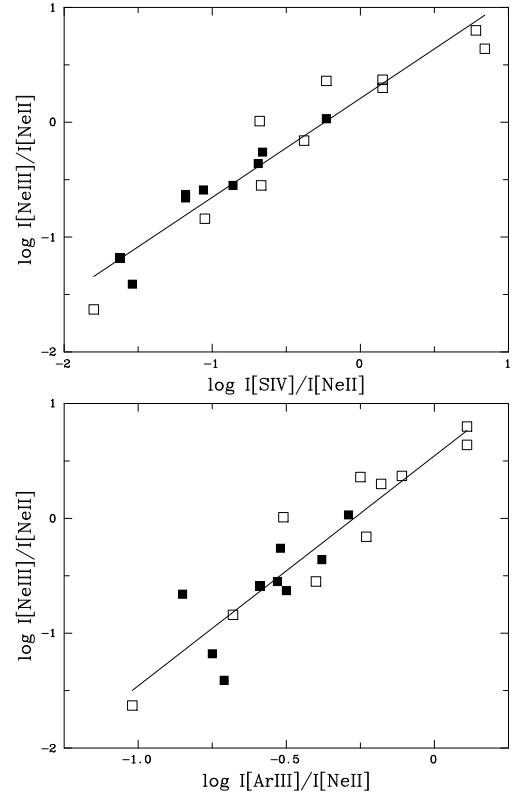}
\caption{(Upper panel) Correlation between $I$[S\,{\sc iv}]/$I$[Ne\,{\sc 
ii}] and $I$[Ne\,{\sc iii}]/$I$[Ne\,{\sc ii}] for H\,{\sc ii} regions
located in the inner (filled squares) and outer (open squares) Milky Way 
using ISO/SWS fluxes 
from Peeters et al. (2002), dereddened as described in the text; (lower
panel) As above except for correlation between $I$[Ar\,{\sc 
iii}]/$I$[Ne\,{\sc ii}] and $I$[Ne\,{\sc iii}]/$I$[Ne\,{\sc ii}].}
\label{midIR}
\end{center}
\end{figure}

The comparison between the predicted [S\,{\sc iv}]/[Ne\,{\sc ii}] ratio 
and empirical results is much poorer than for
the [Ne\,{\sc iii}]/[Ne\,{\sc ii}] ratio. As such, one might conclude that
no  straightforward means of determining temperatures exists from 
ground-based data alone. However, it has been established that there is
a reasonably tight  correlation between the  [S\,{\sc iv}]/[Ne\,{\sc ii}] and [Ne\,{\sc 
iii}]/[Ne\,{\sc  ii}] ratios, as shown by \citet*{Groves08}. For the 
sample of  Milky Way compact H\,{\sc ii} regions observed by 
\cite{Peeters02} we obtain a linear, least-squares fit  
\begin{eqnarray*}
\log I[{\rm Ne\,III}]/I[{\rm Ne\,II}] & = & 0.86 \pm 0.06 (\log I[{\rm 
S\,IV}]/I[{\rm Ne\,II}])\\
   &  &  + 0.21 \pm 0.06
\end{eqnarray*}
after correcting line fluxes for extinction (as discussed above). This is
presented in Fig.~\ref{midIR}. For comparison,
\cite{Groves08} obtained coefficients 0.74 and 0.42 for 65 Galactic H\,{\sc ii} regions from 
\cite{Giveon02}, {\it neglecting} interstellar extinction. 

We use this conversion to apply our empirical [Ne\,{\sc  iii}]/[Ne\,{\sc 
ii}]  temperature calibration to  [S\,{\sc iv}]/[Ne\,{\sc ii}]. This is 
also  presented in Fig.~\ref{U_plot},  from which   similar consistency 
with  empirical results to that achieved  for [Ne\,{\sc iii}]/[Ne\,{\sc 
ii}] is obtained, revealing
\begin{eqnarray*}
(T_{\rm eff}/kK)_{2 Z_{\odot}} & = & +46.79 \pm 1.51 \\
                               &   & +5.39 \pm 1.69
(\log I[{\rm S\,IV}]/I[{\rm Ne\,II}]) 
\end{eqnarray*}
for the super-solar case.
%and 
%\begin{eqnarray*}
%(T_{\rm eff}/kK)_{Z_{\odot}} & = & +40.92 \pm 0.22 \\
%                               &   & +3.44 \pm 0.15
%(\log I[{\rm S\,IV}]/I[{\rm Ne\,II}]) \\
%                               &   & +0.19 \pm 0.16
%(\log I[{\rm S\,IV}]/I[{\rm Ne\,II}])^2 
%\end{eqnarray*}
%for the solar case. 
Again, we include the predicted subtypes of Galactic 
compact H\,{\sc ii} regions from our empirical [S\,{\sc iv}]/[Ne\,{\sc 
ii}] calibrations in Table~\ref{ne3_ne2}. Since [S\,{\sc iv}] is relatively
sensitive to extinction, if we were to use extinctions inferred from their
H--K colours (e.g. $A_{\rm K}$ = 4.9 mag for W31 \#26) rather than the average
of the naked O stars,  we would obtain temperatures up to 3kK higher. It is 
likely that their actual extinctions lie between these two extremes due to the
contribution of hot circumstellar dust. 
%For Tr~14 and NGC~3603, the 
%solar [S\,{\sc iv}]/[Ne\,{\sc ii}] calibration infers temperatures of 
%39--41kK (O5--6 subtypes), in reasonable agreement with their dominant
%stellar content.

\begin{figure}
\includegraphics[width=\columnwidth, clip]{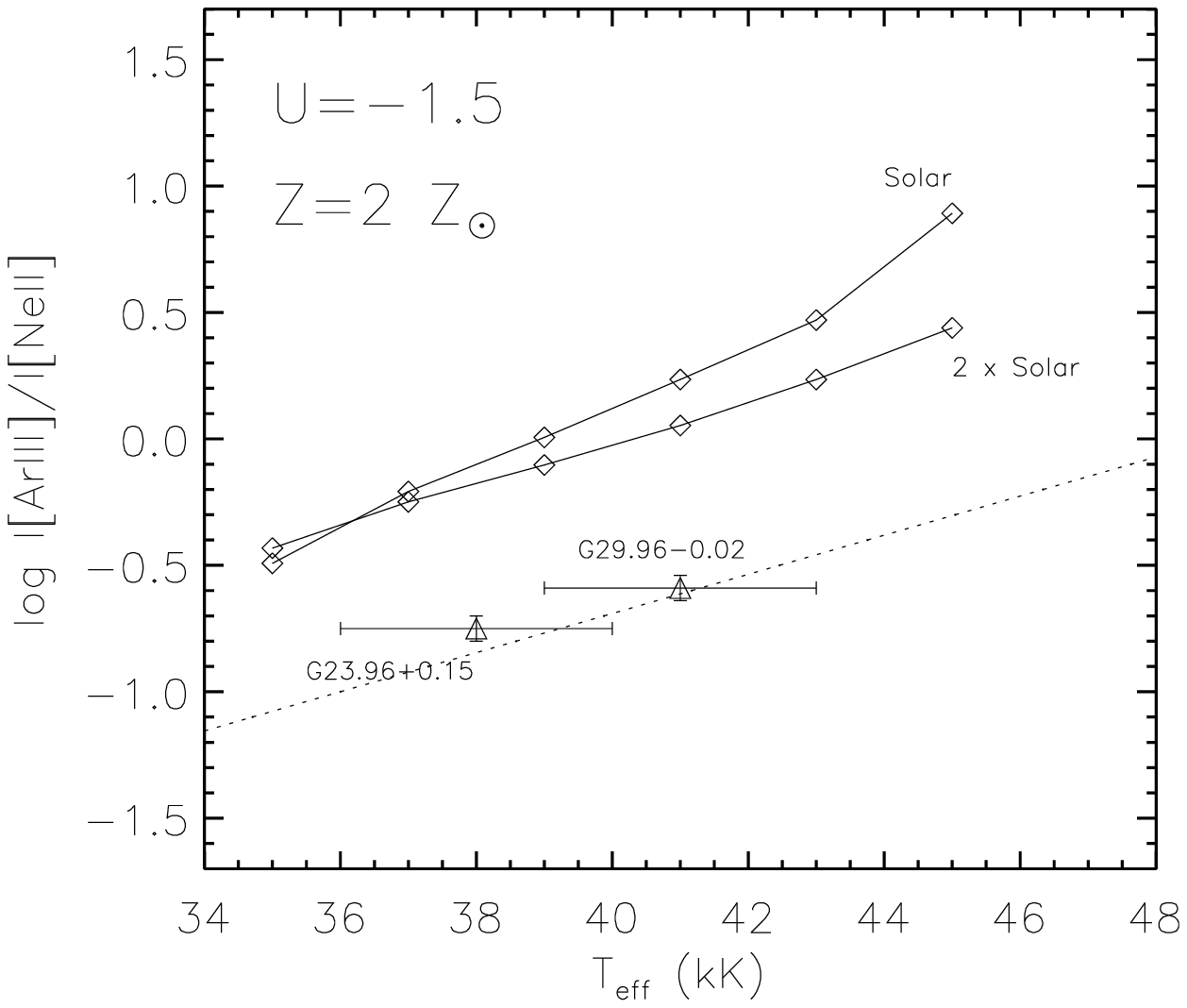} 
\caption{Effective temperature versus log 
[Ar\,{\sc iii}]/[Ne\,{\sc iv}] for our $\log U=-1.5$ solar and 2$\times$ 
Solar models
together with inner Milky Way sources and resulting empirical calibration (dotted-line).
%(Lower panel) As above, for outer Milky Way sources and our empirical
%calibration (dotted-line)
}
\label{U_plot2}
\end{figure}

In principle, either the [Ne\,{\sc iii}]/[Ne\,{\sc ii}] or [S\,{\sc 
iv}]/[Ne\,{\sc ii}] calibration may be applied to H\,{\sc ii} regions 
ionized by early or mid-type O stars, since late-type O stars possess too 
low photon fluxes at the energies required to produce either Ne$^{2+}$ 
($\geq$ 41eV) or S$^{3+}$ ($\geq$ 35eV). Therefore, we have also 
considered the potential role of [Ar\,{\sc iii}] 9.0$\mu$m from the two 
\hii\  regions (this line was not covered by IRS/SH datasets). For the 
\cite{Peeters02} sample of H\,{\sc ii} regions as discussed above, we 
obtained [Ar\,{\sc iii}] line intensities using an extinction correction 
of
\begin{eqnarray*}
A_{\rm 9.0\micron} & = & 0.719\,A_{K}
\end{eqnarray*}
and obtained a similar quality fit to that of I[S\,{\sc iv}]/I[Ne\,{\sc 
ii}] 
above (see Fig.~\ref{midIR}), namely
\begin{eqnarray*}
\log I[{\rm Ne\,III]/I[Ne\,II}] & = & 2.06 \pm 0.21 (\log I[{\rm 
Ar\,III]/I[Ne\,II]}) \\
                               &    & + 0.55 \pm 0.11
\end{eqnarray*}
allowing estimates of temperatures to be made from the [Ar\,{\sc 
iii}]/[Ne\,{\sc ii}] ratio. Unfortunately, this is relatively insensitive 
to temperature and so should only be considered if {\it neither} of the primary 
calibrations are available. Nevertheless, the comparison between 
empirical results for G23.96+0.15 and G29.96--0.02 and our metal-rich 
calibration, presented in Fig.~\ref{U_plot2}, is reasonable. For 
completeness, the super-solar calibration may be expressed as:
\begin{eqnarray*}
(T_{\rm eff}/kK)_{2 Z_{\odot}} & = & +48.92 \pm 1.79 \\
                               &   & +12.92 \pm 3.47
(\log I[{\rm Ar\,III}]/I[{\rm Ne\,II}]) \\
\end{eqnarray*}
%Similar results are 
%obtained for the ONC, G110.10+0.05 to those 
%presented for other diagnostics above, for which the solar calibration is:
%\begin{eqnarray*}
%(T_{\rm eff}/kK)_{Z_{\odot}} & = & +42.31 \pm 0.21 \\
%                               &   & +8.60 \pm 0.46
%(\log I[{\rm Ar\,III}]/I[{\rm Ne\,II}]) \\
%                               &   & +1.11 \pm 0.94
%(\log I[{\rm Ar\,III}]/I[{\rm Ne\,II}])^2 
%\end{eqnarray*}
% However, the inferred temperatures of NGC~3603 and Tr~14 are $\sim$40kK, 
% corresponding to somewhat later subtypes than the observed stellar 
% content.
It should be re-emphasised that these calibrations are solely intended for
compact and ultra-compact H\,{\sc ii} regions for the inner Milky Way,
and are liable to revision once additional empirical results become available.
Nevertheless, it is apparent that the use of any mid-IR nebular diagnostics
to  establish properties of metal-rich O stars will lead to stellar temperatures 
that are too low, by as much as 5\,kK or 10\%.

\section{Discussion} \label{discussion}

\subsection{Massive Star Content of W31}

Let us now return to W31 itself, and specifically its massive star 
content. From Fig.~\ref{emp_fstruc}, the highest ionization stars are 
naked O stars including the highest mass star in W31 \#2, while the lowest
ionization stars (\#1, \#9) are massive YSOs with presumably the lowest 
masses. 

Ideally, one would like to comment upon differences in timescales over 
which circumstellar material is destroyed by young O stars of different 
masses, but the range of masses exhibited by our sample is likely to be 
relatively modest (with the exception of \#2) and the unusual geometry of 
W31 suggests that massive star formation across this cluster was not 
necessarily coeval. The only naked O star from Fig.~\ref{isos} suggesting 
an age spread is \#5. However, as discussed in \S~\ref{distances} it is
plausible that \#5 is a close binary with similar mass components from
which a uniform cluster age would be obtained. 

For the case of a genuinely coeval cluster, it is  likely that stars with the 
highest luminosities -- and hardest extreme UV photon fluxes -- destroy 
circumstellar dust more rapidly than lower luminosity,  softer  extreme UV 
stars, which are conceivably the massive YSOs. From Table~\ref{ne3_ne2}
stellar subtypes of O4--6V are obtained from our [Ne\,{\sc 
iii}]/[Ne\,{\sc ii}] calibration (Table~\ref{ne3_ne2}), i.e. comparable to
the naked O stars. 

W31 \#26, in particular, appears to be in the process of actively clearing
its circumstellar dust, revealing photospheric absorption lines, consistent with a 
subtype of O6V, slightly later than the indirect calibration.  \#15 and \#30, 
on the basis of similar neon ratios to \#26 its mid-IR neon lines likely 
possess similar subtypes, with $\sim$O6--6.5 subtypes for \#1 and \#9. 
The lowest (current) mass  estimated amongst the naked O stars is $\sim 36 
M_{\odot}$ for the O5.5V star \#4 (Table~\ref{stellar}). Consequently, it
is likely that the massive YSO's possess stellar masses of order 30--35 
$M_{\odot}$. Recall also that W31 also hosts a number of
\hii\ regions  \citep{Ghosh}. Such sources are still too deeply embedded to
be detected at near--IR wavelengths, most likely as a result of
still lower luminosities/masses (late O subtypes?). If the \hii\ regions
are genuinely coeval with the naked O stars and massive YSO's in W31, a 
timescale in excess of 0.5 Myr is required for late O stars to sufficiently
clear their environment to be detected at near-IR wavelengths.

\subsection{Analysis of embedded stellar populations from mid-IR fine 
structure lines.}

Our study has focused upon compact and ultra-compact H\,{\sc ii} regions 
within the inner  Milky Way. We find that all mid-IR nebular diagnostics
will lead to stellar temperatures  that are too low, by as much as 5\,kK or 10\%.
However, mid-IR fine structure lines also have application in
external galaxies, ranging from metal-rich H\,{\sc ii} regions in spiral 
galaxies \citep{Rubin07}, starbursts \citep{Thornley00, Verma, Brandl06} 
and ultraluminous infrared galaxies \citep[ULIRGs][]{Lutz98}.

Within this diverse sample, the mid-IR has the greatest diagnostic role 
for the highly obscured cases,  either young, embedded massive clusters 
within metal-poor starbursts  such as NGC\,5253 \citep{Crowther99},
He~2--10 \citep{Vacca02}, or metal-rich ULIRG's \citep{Genzel98}.

Of course, the aim of studies of starbursts is generally to obtain ages 
and/or stellar content (mass, Initial Mass Function) yet similar 
approaches to the present study are usually followed, 
involving photoionization models and  ionizing flux distributions 
from stellar atmosphere models coupled to 
evolutionary predictions through population synthesis codes 
\citep[e.g.][]{Pindao02}.

Problems highlighted here and elsewhere \citep{Morisset02, MartinH02a,
MartinH02b} for H\,{\sc ii} regions in the {\it inner} Milky Way probably 
have  their origins in both (i) incomplete opacities from all relevant 
ions in 
stellar atmosphere models allowing for non-LTE effects and stellar winds; 
(ii) 1D photoionization models. In principle, the second of these can be 
tested against 3D photoionization codes such as MOCASSIN 
\citep*{Ercolano}. 
Until then, similar empirical 
calibrations of photoionization models across a range of 
metallicities may be the best approach.  It may be significant that the 
discrepancy for the Orion Nebula Cluster is significantly less severe than 
for G29.96--0.02 and G23.96+0.15. 
A further test would be for an extensive sample of 
H\,{\sc ii} regions in the outer Milky Way and the Magellanic Cloud, These could 
potentially include 30  Doradus (LMC) and/or NGC~346 (SMC). Beyond the 
Magellanic Clouds, spatial resolution prevents spectroscopy of individual 
stars within compact star clusters.

\section{Summary}\label{summary}

We present near-IR (VLT/ISAAC) and mid-IR ({\it Spitzer}/IRS) spectroscopy 
of 
massive stars within the young Milky Way cluster G10.2--0.3 (W31). Our
main results may be summarised as follows:
\begin{enumerate}

\item $H$- and $K$-band spectroscopy of naked O stars broadly confirms
subtypes from \citet{Blum01} from which a refined cluster
distance (3.3 kpc) and age ($\sim$0.6 Myr) are obtained. 

\item W31 \#26, one of the massive YSOs from \citet{Blum01} is shown to
possess photospheric features, consistent with a subtype of O6V, 
in addition to near-IR circumstellar dust emission. This suggests it is 
in the very process of clearing its immediate environment.

\item Mid-IR fine-structure line ratios of W31 stars overlap with other 
Milky Way and Magellanic Cloud H\,{\sc ii} regions on a [Ne\,{\sc 
iii}]/[Ne\,{\sc ii}] to [S\,{\sc iv}]/[S\,{\sc iii}] diagram.

\item Following \cite{Morisset04}, a comparison of the mid-IR radiation 
hardness parameter, $\eta$(S-Ne) = ([Ne\,{\sc iii}]/[Ne\,{\sc 
ii}])/([S\,{\sc iv}]/[S\,{\sc iii}] versus [Ne\,{\sc iii}]/[Ne\,{\sc ii}] 
allows dependencies upon effective temperature and ionization parameter, 
$U$ to be tested. Predicted solar metallicity (stellar and nebular) models 
differ greatly from empirical $T_{\rm eff}$ and $U$, with small 
improvements in both for 2$\times$ solar grids.

\item Similar studies are planned for metal-poor environments, 
which should establish whether the problem is most severe at high metallicity.
Initial results for the
ONC do show an improved agreement between predictions and observations,
although larger samples are required for statistically robust results.
If the discrepancy were to disappear at lower metallicities
it would suggest that either the extreme UV metal line blanketing of
early-types stars is incomplete, or there is a problem with 
photoionization models at high metallicity.

\item  We show that an empirical correction to the predicted  
[Ne\,{\sc  iii}]/[Ne\,{\sc ii}] ratio against $T_{\rm 
eff}$ for $\log U$=--1.5 and 2$\times Z_{\odot}$ provides  a reasonable 
match to stellar results for W31 and two inner Milky Way \hii\ regions.
Estimates of the O subtypes of the ionizing stars in other 
inner Galactic H\,{\sc ii} regions ($R_{\rm GC} <$ 7~kpc) are obtained.
This approach is only practical for early- and mid- O stars, in view of
the weakness of [Ne\,{\sc iii}] 15.5$\mu$m for late O-types.

\item 
For ground-based datasets lacking [Ne\,{\sc iii}] observations, we have 
used a correlation between [Ne\,{\sc iii}]/[Ne\,{\sc ii}] and [S\,{\sc 
iv}]/[Ne\,{\sc ii}] -- see also \citet{Groves08} -- to provide a 
calibration of [S\,{\sc iv}]/[Ne\,{\sc ii}] against $T_{\rm eff}$ for 
super-solar compact H\,{\sc ii} regions. Since only early- and mid-
O stars provide significant [S\,{\sc iv}] emission, we have 
also obtained a similar relation for 9.0$\mu$m [Ar\,{\sc iii}]/12.8$\mu$m
[Ne\,{\sc ii}]  versus $T_{\rm eff}$,  although this should only be applied when other 
diagnostics are unavailable, with both [Ar\,{\sc iii}] and [Ne\,{\sc ii}] 
expected to be weak in late-type O stars.
\end{enumerate}

Finally, in view of the apparent discrepancy between stellar and 
nebular results at high metallicity compact H\,{\sc ii} regions, 
studies of more straightforward templates are sought, to enable the 
present calibration to be put on a more robust footing. 

\section*{Acknowledgements}
JPF would like to acknowledge financial support from STFC,
CLB acknowledges financial support from FAPESP and 
PSC thanks the NSF for continuous support. We wish to thank John Hillier
and Gary Ferland for maintaining CMFGEN and CLOUDY.
Specific support for this work 
was partly provided by NASA through an award issued by JPL/Caltech.
We appreciate many useful suggestions from an anonymous referee.

\label{lastpage}

\bibliographystyle{mn2e}
\bibliography{bib_list}

\end{document}